\documentclass[aps,pra,twocolumn,superscriptaddress,notitlepage,amsmath,amssymb,floatfix]{revtex4-2}

\usepackage{amsmath,amsthm,amssymb,bm,graphicx,float,xcolor,changes,mathtools,enumitem}
\usepackage[colorlinks=true,urlcolor=blue,linkcolor=blue,citecolor=green,bookmarks=false]{hyperref}

\allowdisplaybreaks

\newcommand{\be}{\begin{equation}}
\newcommand{\ee}{\end{equation}}
\newcommand{\6}{\partial}

\newcommand{\bj}{\boldsymbol{j}}

\newcommand{\bx}{\boldsymbol{x}}
\newcommand{\bz}{\boldsymbol{z}}

\begin{document}

\title{Universal properties and dynamical bosonization of strongly interacting one-dimensional  anyons}

\begin{abstract}

We study a one-dimensional system of interacting anyons with short-range interactions under external confinement. This model, which describes 
anyons with zero-range $s$-wave and $p$-wave scattering interactions in the strong interaction regime continuously interpolates between 
spin-polarized fermions with $p$-wave interactions and free bosons. At zero temperature, 
the correlation functions decay exponentially with distance, with oscillations governed by the statistics parameter. The decay rate is maximal for 
$p$-wave fermions and decreases monotonically as the statistics parameter approaches the bosonic limit, where it vanishes.
The momentum distribution is asymmetric, a hallmark of one-dimensional anyons, and takes the form of a shifted Lorentzian with universal power-law 
tails, $\lim_{k \to \pm \infty} n(k)\sim C/k^2$. We prove analytically that, following release from a harmonic trap, the asymptotic momentum 
distribution converges to that of free bosons in the same trap, a phenomenon known as dynamical bosonization.
We also establish the universality of the groundstate $n$-particle reduced density matrices: their natural occupations are independent of the 
confining potential, while the associated natural $n$-functions for different confinements are related through a simple analytical transformation. 
In particular, for the one-particle reduced density matrix, we derive exact expressions for both the natural occupations and the natural orbitals 
at arbitrary particle number. These results extend and unify earlier partial findings for $p$-wave fermions, and they provide a clear conceptual 
explanation of the double degeneracy observed in their spectrum.
 
\end{abstract}

\author{Ovidiu I. P\^{a}\c{t}u}
\affiliation{Institute for Space Sciences, Bucharest-M\u{a}gurele, R 077125, Romania}

\maketitle

\section{Introduction}\label{s1}

One-dimensional (1D) anyons are particles with fractional exchange statistics, whose wavefunctions acquire a non-trivial phase factor when two 
particles exchange positions. Compared with the more familiar two-dimensional (2D) case \cite{LM77,Wilc82}, realizing fractional statistics in 
one dimension is more subtle. In 1D, particles can exchange positions only through collisions, which requires a convention for assigning the 
sign of the statistical phase to each pair \cite{AN07}. This broken spatial invariance leads to an asymmetric momentum distribution, a 
distinctive feature of one-dimensional anyons.
From a theoretical perspective, both equilibrium and nonequilibrium properties of 1D anyons have been investigated extensively over the past two 
decades, in both continuum \cite{Kund99,BGO06,Gira06,SSC07,CM07,PKA07,PKA08a,SC08,delC08,BGK08,BFGL08,HZC08,CS09,Grei09,BS12,WRDK14,Zinn15,MPC16,
Hao16,PC17,RCCM19,SPC20,Patu20,HK20,MD21,BJEP21,HK22,Patu23b,MYC23,VVGD25}  and lattice settings \cite{AN07,HZC09,BCM09,KLMR11,LD12,RFB14,WWZ14,
TEP15,GS15,AFS16,ZGFS17,LGDG18,MWS20,GIZ21,TGB21,ZNIG22,CG22,Wang22,NWKE24,TLBI24,PKF24,BJHP25}. More recently, fractional statistics have been realized 
experimentally across a variety of physical platforms \cite{ZYWD22,KSLK24,DWHV25}.

In the absence of a spin-statistics theorem in one dimension, these systems exhibit distinctive properties and regimes not found in higher 
dimensions.  A well-known example is the Bose-Fermi mapping \cite{Gira60}, originally established between hard-core bosons (the Tonks-Girardeau 
gas) and free fermions, and later extended to finite interactions \cite{CS99}. In its most general form, the mapping relates the wavefunctions of 
the integrable Lieb-Liniger model of bosons \cite{LL63} with $\delta$-function ($s$-wave) interactions to those of spin-polarized fermions with 
$\delta'(x)$-function ($p$-wave) interactions \cite{CS98,CS99,GO04,GB04,BEG05,GM06,HZC07,Pric08,ILGG10,CLH16,Cui16,YGC16,PCC18,STN18,XGZS18,KS20,
KS23,SRAJ24,SRAJ25}. The coupling constants play dual roles: weak repulsive interactions in the bosonic system correspond to strong attraction in 
the fermionic system, and vice versa. More recently, this mapping has been generalized to particles with arbitrary exchange statistics and at 
arbitrary coupling strength \cite{WCC25,HBB25a,HBB25b} (see also \cite{Gira06,PKA08a} for the case of impenetrable particles). This extension 
establishes a correspondence between Lieb-Liniger Bethe ansatz eigenstates (or, equivalently, $p$-wave fermionic eigenstates) and anyonic 
wavefunctions, which are eigenstates of a Hamiltonian featuring both $s$-wave and $p$-wave interactions with identical scattering lengths. In this 
work, we investigate the properties of these particles, hereafter referred to as 
$sp$-anyons, in the strongly attractive regime, where the many-body wavefunction can be expressed as the product 
of the wavefunction of trapped free bosons and an anyonic generalization of Girardeau’s Bose-Fermi mapping \cite{Gira60}.

This system may be viewed as the $p$-wave analogue of the well-known  anyonic Tonks-Girardeau gas \cite{HZC08,PKA08a,CS09,WRDK14,MPC16,
Hao16}, whose wavefunction consists of the Slater determinant of trapped free fermions multiplied by the anyonic Bose-Fermi mapping. In the anyonic Tonks-Girardeau
case, diagonal correlations coincide with those of free fermions, while in our case they coincide with those of free bosons. By contrast, 
the off-diagonal correlations interpolate smoothly between those of $p$-wave fermions and free bosons as the statistical parameter is varied. This 
is captured by the one-particle reduced density matrix (the field-field correlator), which in a homogeneous system at zero temperature decays 
exponentially with distance, modulated by oscillations whose frequency depends on the exchange statistics. The decay rate is maximal for $p$-wave 
fermions \cite{BEG05} and decreases monotonically with the statistics parameter, vanishing in the bosonic limit.
The oscillatory modulation of the correlations leads to an asymmetric momentum distribution, obtained as the Fourier transform of the one-particle 
density matrix. This distribution takes the analytic form of a shifted Lorentzian with universal $1/k^2$ power-law tails, 
characteristic of systems with $p$-wave interactions \cite{BEG05,YTZ15,YU15,LTSY16,HZCZ16,PLH16,QCY16,Cui16,ZHZ17,STN18,Qin18,HZ21,MLZ24}. 
The asymmetry, rooted in the broken spatial invariance of the wavefunction, is a defining 
feature of 1D anyons and  provides a clear experimental signature of fractional statistics.

The nonequilibrium dynamics of $sp$-anyons displays several distinctive features. We prove analytically and confirm numerically that, after 
release from a harmonic trap, the asymptotic momentum distribution acquires the same form as the density profile of free bosons in the trap, 
and coincides with the momentum distribution of free bosons in the initial confinement. This phenomenon, known as \emph{dynamical bosonization} 
\cite{GM06,delC08}, contrasts with the case of  Tonks-Girardeau anyons, where the asymptotic quantities are instead related to those of free fermions 
\cite{RM05,MG05,delC08,ASYP21,Patu23}. Furthermore, a sudden quench of the trap frequency from $\omega_0$ to $\omega_1$ induces breathing 
oscillations \cite{FCJB14,WMLZ20}. However, unlike Tonks-Girardeau anyons, whose momentum distribution exhibits an additional narrowing at $\omega_1 t = 
(m + 1/2)\pi,\, m=0,1,\dots$ \cite{ABGK17,Patu20}, this feature is absent in the $p$-wave case.

Recent work \cite{KS23} has shown that, due to the particular structure of the groundstate wavefunction, the $n$-particle reduced density 
matrices (RDMs) of $p$-wave fermions display a universal form: their natural occupations are independent of the external potential, while the 
associated natural $n$-point functions for different confinements are related by a simple transformation. We demonstrate that this universality 
extends to $sp$-anyons. Furthermore, we derive and solve the ordinary differential equation governing the natural orbitals under the 
appropriate boundary conditions, thereby obtaining both the natural occupations (eigenvalues) and natural orbitals analytically. Our results 
generalize the partial findings of \cite{KS23,SRAJ24} for $p$-wave fermions and, moreover, provide an explicit analytical explanation for the 
double degeneracy observed in their spectrum.

The structure of the paper is as follows. In Sec.~\ref{s2} we introduce the model and its wavefunctions. Section~\ref{s3} analyzes the momentum 
distributions for both homogeneous and trapped systems. The nonequilibrium dynamics, focusing on free expansion and breathing oscillations, is 
investigated in Sec.~\ref{s4}. In Sec.~\ref{s5} we present the universal form of the eigenvalue problem for RDMs, and in Sec.~\ref{s6} we provide 
the complete solution for the one-particle RDM. Additional technical details are collected in two appendices.

\section{One-dimensional anyons with zero-range $s$-wave and $p$-wave scattering interactions}\label{s2}

\subsection{Hamiltonian and wavefunctions}\label{s21}

In this paper, we investigate the properties of a general class of one-dimensional anyons with short-range interactions, introduced in 
\cite{WCC25,HBB25a,HBB25b}. Before presenting the Hamiltonian and its eigenstates, it is useful to recall some of the fundamental properties of 
fermionic and bosonic systems with short-range interactions in one dimension.

We begin with spin-polarized fermions interacting via $p$-wave scattering \cite{CS98,CS99,GO04,GB04,BEG05,GM06,HZC07,Pric08,ILGG10,CLH16,Cui16,YGC16,PCC18,
STN18,XGZS18,KS20,KS23,SRAJ24,SRAJ25}. A system of $N$ trapped fermions with $p$-wave interactions is described by the Hamiltonian
\be\label{defhamm}
H_-=T+\sum_{j=1}^N V(z_j)+\sum_{1\le j<k\le N}V_-(z_j-z_k)\, ,
\ee
where $T=-\frac{\hbar^2}{2m}\sum_{j=1}^N\frac{\6^2}{\6 z_j^2}$ is the kinetic energy operator, $V(z)$ is the external trapping potential and the 
$p$-wave interaction [$\delta'(z)$ potential] is formally described by 
\be\label{defvm}
V_-(z)=g_-\frac{\overleftarrow{\6}}{\6 z}\delta(z)\frac{\overrightarrow{\6}}{\6 z}\, ,\ \ g_-=\frac{2\hbar^2}{m} a_-\, , 
\ee
with $m$ the particle mass and $a_-$ is the scattering length. The eigenfunctions of the Hamiltonian \eqref{defhamm}, denoted $\psi_-(\bz_N)$ with 
$\bz_N=(z_1,\dots,z_N)$, are antisymmetric
\be
\psi_-(\cdots,z_i,z_{i+1},\cdots)=-\psi_-(\cdots,z_{i+1},z_{i},\cdots)\, 
\ee
and exhibit a discontinuity at particle coincidences while their derivatives remain continuous \cite{CS99}.

In contrast, bosons in one dimension interact through the usual $\delta$-function potential. A system of $N$ bosons with short-range interactions is 
described by
\be\label{defhamp}
H_+=T+\sum_{j=1}^N V(z_j)+\sum_{1\le j<k\le N}V_+(z_j-z_k)\, ,
\ee
where $T$ and $V(z)$ are as above and 
\be\label{defvp}
V_+(z)=g_+\delta(z)\, ,\ \ g_+=-\frac{2\hbar^2}{m} \frac{1}{a_+}\, , 
\ee
with $a_+$ the scattering length. In the absence of external confinement, the Hamiltonian \eqref{defhamp} reduces to the integrable Lieb-Liniger model 
\cite{LL63}. The eigenfunctions $\psi_+(\bz_N)$   are symmetric 
\be
\psi_+(\cdots,z_i,z_{i+1},\cdots)=\psi_+(\cdots,z_{i+1},z_{i},\cdots)\, ,
\ee
continuous at particle coincidences, but their derivatives are discontinuous \cite{LL63}.

When the scattering lengths coincide, $a_+=a_-$ (equivalently, $g_+ g_- = -4\hbar^4/m^2$), the eigenfunctions of (\ref{defhamm}) and (\ref{defhamp}) 
share the same energy spectrum and are related via Girardeau’s Bose-Fermi mapping \cite{CS99,Gira60}:
\be
\psi_-(\bz_N)=A_G(\bz_N)\psi_+(\bz_N)\, ,
\ee
where 
\be\label{bfmapping}
A_G(\bz_N)=\prod_{1\le j<k\le N}\epsilon(z_j-z_k)\, ,
\ee
and $\epsilon(z)$ is the sign function: $\epsilon(z)=1$ for $z>0$, $\epsilon(z)=-1$ for $z<0$, and undefined at $z=0$.

Our system of interest consists of anyons with zero-range $s$-wave and $p$-wave scattering interactions ($sp$-anyons). 
The corresponding eigenfunctions are given by
\be\label{wavea}
\psi(\bz_N)=A_\kappa(\bz_N)\psi_+(\bz_N)\, ,
\ee
where
\be\label{mapping}
A_\kappa(\bz_N)=\prod_{1\le j<k\le N} e^{i\frac{\pi\kappa}{2}\epsilon(z_j-z_k)}\, ,
\ee
is the anyonic generalization of the Bose–Fermi mapping~\cite{Gira06,PKA08a}. Here, $\kappa$ denotes the statistical parameter, which we restrict 
to $\kappa \in [0,1]$. The wavefunctions in Eq.~\eqref{wavea} interpolate continuously between the $p$-wave fermionic case at $\kappa = 1$ and 
the Lieb–Liniger bosonic case at $\kappa = 0$.
The eigenfunctions in Eq.~\eqref{wavea} correspond to the Hamiltonian~\cite{WCC25,HBB25a,HBB25b}
\be\label{defhama}
H=T+\sum_{j=1}^N V(z_j)+\sum_{j<k} V_-(z_j-z_k)+\sum_{j<k} V_+(z_j-z_k)\, ,
\ee
where the interaction potentials $V_\pm$ are characterized by equal scattering lengths ($a \equiv a_+ = a_-$). Note that $V_- \psi_+(\bz_N) = 
V_+ \psi_-(\bz_N) = 0$~\cite{CS98,CS99}.

The states defined in Eq.~\eqref{wavea}, which we refer to as $sp$-anyons, were introduced by Hidalgo-Sacoto, Busch, and Blume~\cite{HBB25a,HBB25b} 
(see also Ref.~\cite{WCC25}), where they were termed bosonic anyons. Similarly, wavefunctions of the form $\psi(\bz_N) = A_\kappa(\bz_N)
\psi_-(\bz_N)$ were referred to as fermionic anyons in Refs.~\cite{HBB25a,HBB25b}. Although the states in Eq.~\eqref{wavea} share the same 
energy spectrum as $\psi_\pm(\bz_N)$, they exhibit anyonic exchange symmetry,
\be
\psi(\cdots,z_i,z_{i+1},\cdots)=e^{i\pi \kappa \epsilon(z_i-z_{i+1})}\psi(\cdots,z_{i+1},z_i,\cdots) .
\ee

The two-body zero-range interactions impose the following asymptotic behavior on the wavefunctions when the interparticle 
distance $z_{jk} = z_j - z_k$ approaches zero while all other coordinates are held fixed~\cite{WCC25,HBB25a,HBB25b}:
\be\label{shortbehav}
\psi(\bz_N)\ \underset{z_{jk}\rightarrow 0}{\longrightarrow} \ \, h^{(2)}(z_{jk})\Phi(Z_{jk},\{z_l\}_{l\ne j,k})\, ,
\ee
where
\be
h^{(2)}(z_{jk})=\cos\left(\frac{\pi\kappa}{2}\right)h_+^{(2)}(z_{jk})+ i \sin\left(\frac{\pi\kappa}{2}\right)h_-^{(2)}(z_{jk})\, ,
\ee
and
\begin{align}
h_+^{(2)}(z_{jk})&=1-|z_{jk}|/a\, ,\label{shorts}\\
h_-^{(2)}(z_{jk})&=\mathrm{sign}(z_{jk})-z_{jk}/a\, \label{shortp}\, .
\end{align}
In Eq.~\eqref{shortbehav}, $\Phi(Z_{jk}, \{z_l\}_{l \ne j,k})$ denotes the part of the wavefunction that is independent of $z_{jk}$, while
$Z_{jk} = (z_j + z_k)/2$ is the two-body center-of-mass coordinate.
Equation~\eqref{shortbehav} reveals that the short-range behavior of the anyonic wavefunction is a superposition of the $s$-wave and 
$p$-wave short-distance components, described by Eqs.~\eqref{shorts} and~\eqref{shortp}, respectively.
The discontinuity proportional to $|z_{jk}|/a$ in Eq.~\eqref{shorts}, induced by the $s$-wave potential, leads to a $1/k^2$ decay of the 
Fourier transform of an arbitrary wavefunction at large $k$~\cite{OD03,STN18}. Similarly, the $\mathrm{sign}(z_{jk})$ discontinuity in 
Eq.~\eqref{shortp}, originating from the $p$-wave potential, results in a $1/k$ decay of the Fourier transform at large $k$~\cite{Cui16,STN18,XGZS18}.  
A detailed analysis performed in Refs.~\cite{HBB25a,HBB25b} shows that the high-momentum tail of the momentum distribution of $sp$-anyons, 
defined in Eq.~\eqref{defnk}, is given by
\begin{align}\label{largekmom}
\lim_{|k|\rightarrow \infty}n(k)=&\frac{4 C_2}{k^2}\sin^2\left(\frac{\pi\kappa}{2}\right)+\frac{4C_2}{ak^3}\sin(\pi\kappa)\nonumber\\
&\ +\frac{8C_3}{k^3}\sin^2\left(\frac{\pi\kappa}{2}\right)\sin(\pi\kappa)+\mathcal{O}(k^{-4})\, ,
\end{align}
where $C_2$ and $C_3$ are Tan’s contacts~\cite{Tan08b,BZ11,PK17}, defined in terms of the two- and three-particle reduced density matrices.
Two remarks are in order.  
First, the leading contribution to the momentum tail, which scales as $1/k^2$, originates from $p$-wave scattering, whereas $s$-wave scattering contributes only at order $1/k^4$~\cite{Tan08b,BZ11,PK17}.  
Second, it is important to emphasize that the large-momentum behavior described by Eq.~\eqref{largekmom} is \emph{universal}: it holds for any particle number, interaction strength, or external potential, both at zero and finite temperature, and remains valid in both equilibrium and nonequilibrium situations like those induced by variations in the external potential.  
The Tan contacts, however, are system-dependent and therefore non-universal.

In this work, we focus on the regime of strong attraction, $a=-\infty$, where $V_+(z)=0$ and $\psi_+(\bz_N)$ reduces to the free-boson wavefunction. 
In the groundstate, the anyonic wavefunction becomes 
\be\label{wavegs}
\psi(\bz_N)=A_\kappa(\bz_N)\psi_+^{(0)}(\bz_N)\, , \  \psi_+^{(0)}(\bz_N)=\prod_{j=1}^N \phi_0(z_j)\, ,
\ee
with $\phi_0(z)$ denoting the lowest-energy single-particle orbital. 
In Sec.~\ref{s4}, we also investigate the nonequilibrium dynamics induced by a sudden change in the trapping potential. 
The Bose–Fermi mapping and its anyonic generalization, Eq.~\eqref{mapping}, remain valid in the presence of time-dependent 
external potentials~\cite{GW00,MG05,GM06}, with the time-dependent many-body wavefunction given by
\be\label{wavegstime}
\psi(\bz_N,t)=A_\kappa(\bz_N)\psi_+^{(0)}(\bz_N,t)\, , \  \psi_+^{(0)}(\bz_N,t)=\prod_{j=1}^N \phi_0(z_j,t)\, ,
\ee
where $\phi_0(z,t)$ is the time-dependent single-particle orbital. 
In both equilibrium and nonequilibrium situations, the short-range behavior of the wavefunction~\eqref{wavegstime} is governed by 
Eq.~\eqref{shortp}; that is, it exhibits a $\mathrm{sign}(z_{jk})$ discontinuity originating from the $p$-wave interaction. As a result, 
the momentum distribution features a universal high-momentum tail of the form $n(k) \sim C_2/k^2$~\cite{Cui16,STN18,XGZS18}.
Throughout the remainder of this work, we set $\hbar=1$.

\subsection{Reduced density matrices}

Important information about a many-body system is encoded in a set of $2n$-point correlation functions, known as the $n$-particle reduced 
density matrices (RDMs). Studying these correlation functions allows one to determine the density profile, momentum distribution, density 
correlations, and to identify off-diagonal long-range order, among other properties.
Introducing 
the notations $\bx_n=(x_1,\cdots,x_n)$, $\bx_n'=(x_1',\cdots,x_n')$,  $\bz_{N-n}=(z_1,\cdots,z_{N-n})$ and  $\bz_{N-n}'=(z_1',\cdots,z_{N-n}')$
the $n$-particle RDM for a state described by $\psi(\bz_N|\bj)$ is defined as (the asterisk denotes complex conjugation)
\begin{align}\label{defrdm}
\rho^{(n)}(\bx,\bx'|\bj)=\frac{N!}{(N-n)!}\int& d\bz_{N-n}\,\psi^*(\bz_{N-n},\bx_n'|\bj) \nonumber\\
&\ \times\psi(\bz_{N-n},\bx_n|\bj)\, .
\end{align}
Here, $\boldsymbol{j}=(j_1,\cdots,j_N)$ is a multi-index specifying the particular state (for example, the groundstate corresponds to 
$\boldsymbol{j}=(0,\cdots,0)$).
Note that some definitions in the literature differ by complex conjugation. Additionally, the relative positions of $\bx_n$ and 
$\bx_n'$ are particularly significant for anyonic correlators \cite{PKA08a}.

\section{Momentum distribution in the groundstate}\label{s3}

The momentum distribution is one of the most important observables characterizing a many-body system which is also directly accessible in 
experiments. Analytically, it is obtained from the Fourier transform of the one-particle reduced density matrix, also known as the 
field-field correlation function.
From the general definition in Eq.~(\ref{defrdm}), the one-particle RDM in a state $\boldsymbol{j}$ is given by
\begin{align}\label{def1rdm}
\rho^{(1)}(x,x'|\boldsymbol{j}) &= N \int dz_1 \cdots dz_{N-1}, \psi^*(z_1,\dots,z_{N-1},x'|\boldsymbol{j}) \nonumber\\
&\qquad \times \psi(z_1,\dots,z_{N-1},x|\boldsymbol{j})\, .
\end{align}
When $x=x'$, this reduces to the real-space density, which satisfies the normalization condition $\int \rho^{(1)}(x,x)\, dx=N$. For  
$sp$-anyons, the real space density coincides with that of the corresponding bosonic state $\psi_+(\bz_N|\boldsymbol{j})$. By contrast, the momentum 
distribution, defined as
\begin{equation}\label{defnk}
n(k) = \frac{1}{2\pi} \int \int e^{-i k(x-x')} \rho^{(1)}(x,x') dx dx'\, ,
\end{equation}
and normalized by $\int n(k)\, dk=N$, differs markedly from the bosonic case. Its most striking feature is its asymmetry with respect to $k$.
For the groundstate, Eqs.~(\ref{wavegs}) and (\ref{def1rdm}) yield
\begin{widetext}
\begin{align}\label{s3:e1}
\rho^{(1)}(x,x')&=N\phi_0^*(x')\phi_0(x)\left(\int_{L_-}^{L_+}\phi_0^*(z)\phi_0(z)e^{-i\frac{\pi\kappa}{2}\left[\epsilon(x-z)-\epsilon(x'-z)\right]}\, 
dz\right)^{N-1}\, ,\nonumber\\
\rho^{(1)}(x,x')&=N\phi_0^*(x')\phi_0(x)\left(1-\gamma(x,x')\int_{x'}^{x}\phi_0^*(z)\phi_0(z)\, dz\right)^{N-1}\, ,
\end{align}
\end{widetext}
where we have used $\int_{L_-}^{L_+} \phi_0^*(z)\phi_0(z)\, dz=1$ and defined
\be\label{defgamma}
\gamma(x,x')=\left[1-e^{-i\pi\kappa\epsilon(x-x')}\right]\epsilon(x-x')\, .
\ee
Here $L_\pm$ denote the system boundaries. For a power-law confining potential, $L_\pm = \pm \infty$, while for hard-wall or periodic boundary 
conditions $L_\pm =\pm L/2$.

\begin{figure*}
\includegraphics[width=1\linewidth]{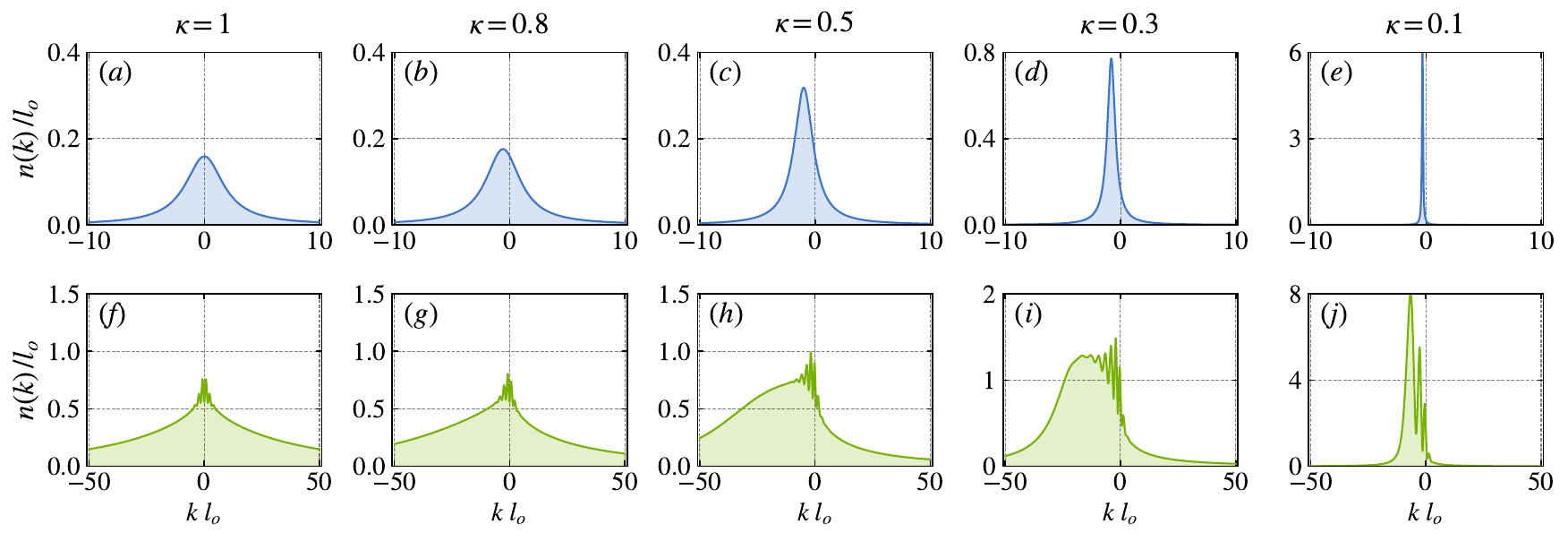}
\caption{
Groundstate momentum distribution of $sp$-anyons for different values of the statistical parameter $\kappa$.  
Results are shown for a homogeneous system in the thermodynamic limit (top row) and for a harmonically trapped system 
with parameters $N=50$, $l_0=m=\omega=1$ (bottom row).  
Note the asymmetry of the momentum distribution, $n(-k)\neq n(k)$, which is a hallmark of one-dimensional anyons for $\kappa \neq 1$.
}
\label{fig1}
\end{figure*}

From this general form of the one-particle RDM one can directly demonstrate a key property of one-dimensional anyons: the asymmetry of the momentum distribution 
\cite{SSC07,SC08,HZC08,HZC09,TEP15}. 
Specifically, Eqs.~(\ref{s3:e1}) and (\ref{def1rdm}) imply that exchanging $x$ and $x'$ is equivalent to complex conjugation:
\be
\rho^{(1)}(x,x')=\left[\rho^{(1)}(x',x)\right]^*\, .
\ee
This contrasts with the bosonic ($\kappa=0$) and fermionic ($\kappa=1$) limits, where $\rho^{(1)}(x,x')$ is purely real and invariant under 
$x \leftrightarrow x'$.
To make this feature explicit, let us consider a translationally invariant system (the argument extends to the general case, albeit with 
additional technical steps), for which $\rho^{(1)}(x,x')$ depends only on the relative coordinate, $\rho^{(1)}(x,x')=\rho^{(1)}(|x-x'|)=
\rho^{(1)}(y)$. The momentum distribution is then given by
\begin{align}
n(k) &= \frac{1}{2\pi}\int_{-\infty}^\infty e^{-i k y} \rho^{(1)}(y)\, dy\, , \nonumber\\
&= \frac{1}{2\pi}\left(\int_{0}^\infty e^{i k y}\,[\rho^{(1)}(y)]^*\, dy + \right. \nonumber\\
&\qquad\qquad\left. +\int_{0}^\infty e^{-i k y}\, \rho^{(1)}(y)\, dy\right)\, ,\nonumber\\
&= \frac{1}{2\pi}\left( \int_0^\infty 2\cos(ky)\, \Re[\rho^{(1)}(y)]\, dy \right. \nonumber\\
&\qquad\qquad\left. + \int_0^\infty 2\sin(ky)\, \Im[\rho^{(1)}(y)]\, dy \right)\, .
\end{align}
When $\kappa=0$ or $\kappa=1$, the RDM is purely real, i.e. $\Im[\rho^{(1)}(y)]=0$, and the momentum distribution is symmetric, $n(k)=n(-k)$. 
For intermediate values of $\kappa$, however, $\Im[\rho^{(1)}(y)] \neq 0$, which introduces a sine contribution and leads to an asymmetric 
momentum distribution, $n(k)\neq n(-k)$. This asymmetry in $n(k)$ is therefore a direct and distinctive manifestation of fractional statistics 
in one dimension, providing a clear physical signature that differentiates anyons from both bosons and fermions.

In the following sections, we will investigate the momentum distribution in two physically relevant settings: (i) a system defined on a ring, 
and (ii) a system  subject to a harmonic trapping potential.

\subsection{System on a ring}

For a system on a ring of circumference $L$, the lowest single-particle orbital is $\phi_0(z)=1/\sqrt{L}$. The overlap appearing in Eq.~(\ref{s3:e1}) 
can then be evaluated explicitly as $\int_{x'}^x\phi^*_0(z)\phi_0(z)\, dz=(x-x')/L.$ Accordingly, the one-particle RDM for a finite ring reads
\begin{align}
\rho^{(1)}(x,x')=\frac{N}{L}\phi_0^*(x')\phi_0(x)\left(1-\gamma(x,x')\frac{(x-x')}{L}\right)^{N-1}\, ,
\end{align}
where $\gamma(x,x')$ is defined in Eq.~(\ref{defgamma}). In the thermodynamic limit ($L,N\to\infty$ with $n=\lim_{L\to\infty}N/L$ fixed), we use 
$\lim_{L\to\infty}(1+x/L)^L=e^x$ to obtain
\begin{align}\label{s4:exp}
\rho^{(1)}(x,x')=n\,e^{-n[1-\cos(\pi\kappa)]|x-x'|} e^{- i n\sin (\pi \kappa) (x-x')}\, .
\end{align}
As in the case of $p$-wave fermions studied in Ref.~\cite{BEG05}, the one-particle RDM decays exponentially with spatial separation. The decay rate is 
maximal for $\kappa=1$ ($p$-wave fermions) and decreases monotonically with decreasing $\kappa$, vanishing in the bosonic limit $\kappa=0$. In addition, 
the anyonic RDM acquires an oscillatory phase factor with frequency set by $\kappa$, a distinctive feature of fractional statistics \cite{SSC07,CM07,PKA07,SC08}.
In the thermodynamic limit (the system is translationally invariant) the momentum distribution (\ref{defnk}) follows from the identity 
$\frac{1}{2\pi}\int e^{i kz}e^{-a |z|}\, dz=\frac{1}{\pi}\frac{a}{a^2+k^2}$ for $a>0$ leading to
\be
n(k)=\frac{1}{\pi}\frac{n[1-\cos (\pi\kappa)]}{n^2[1-\cos(\pi\kappa)]^2+[k +n\sin(\pi\kappa)]^2}\, .
\ee
Thus, the momentum distribution has the form of a shifted Lorentzian. For $\kappa=1$ ($p$-wave fermions) \cite{BEG05} it is centered at $k=0$, while for 
$\kappa=0$ (free bosons) it collapses to a $\delta$-function at $k=0$, as expected for a condensate in the groundstate. For intermediate values $0<\kappa<1$, 
the peak is shifted away from zero momentum, directly reflecting the asymmetry $n(k)\neq n(-k)$ and thereby providing a clear momentum-space signature 
of anyonic statistics. 
At large $|k|$, the momentum distribution exhibits universal power-law tails
\be\label{tanhom}
\lim_{k\rightarrow\pm \infty} n(k)=C/k^2\, ,\ \ C=n[1-\cos (\pi\kappa)]/\pi\, ,
\ee
where $C$ is the Tan contact.
This $1/k^2$ behavior is characteristic of $p$-wave fermions  \cite{BEG05,YTZ15,YU15,LTSY16,HZCZ16,PLH16,QCY16,Cui16,ZHZ17,STN18,Qin18,HZ21,MLZ24} and originates from the 
short-distance structure of the wavefunction (\ref{wavea}). 
At small relative separations, the wavefunction can be expanded as a superposition of $s$-wave and $p$-wave contributions, with the latter providing the dominant 
leading-order contribution \cite{HBB25a,HBB25b} (see also the discussion of the short-distance behavior in  Sec.~ \ref{s21}).
Plots of the momentum distribution for different values of $\kappa$, illustrating the characteristic asymmetry, are shown in the first row of Fig.~\ref{fig1}.

\subsection{Harmonically trapped system}

The overwhelming majority of experiments with ultracold gases are performed in inhomogeneous settings, where the system is confined by an external 
potential that can, to a very good approximation, be taken as harmonic $V(z)=m\omega^2z^2/2$. In the presence of harmonic confinement, translational 
invariance is lost, and the lowest orbital corresponds to the first Hermite function,
\be\label{defphi0}
\phi_0(z)=\frac{1}{l_o^{1/2}\pi^{1/4} }e^{-z^2/(2 l_o^2)}\, ,
\ee
where $l_o=(1/m\omega)^{1/2}$ denotes the harmonic oscillator length. The diagonal part of the one-body density matrix, i.e. the density profile $\rho^{(1)}(x)\equiv\rho^{(1)}(x,x)$,
coincides with that of a system of noninteracting bosons in the same external potential, denoted $\rho_B^{(1)}(x)$:
\be\label{densb}
\rho^{(1)}(x)=\rho^{(1)}_B(x)=\frac{N}{l_o\pi^{1/2} }e^{-x^2/(l_o^2)}\, .
\ee
The wavefunction overlap appearing in Eq.~(\ref{s3:e1}) can be evaluated using the error function, defined by  $\mbox{Erf}(z)=\frac{2}{\sqrt{\pi}}\int_0^z e^{-t^2}\, dz$ which yields 
\be\label{defoverlap}
O(x,x')\equiv\int_{x'}^x \phi^*_0(z)\phi_0(z)\, dz=\frac{1}{2}\left[\mbox{Erf}\left(\frac{x}{l_o}\right)-\mbox{Erf}\left(\frac{x'}{l_o}\right)\right]\, .
\ee
The one-body RDM then takes the form
\begin{align}
\rho^{(1)}(x,x')=\frac{N}{l_o\sqrt{\pi}}e^{-\frac{[x^2+(x')^2]}{2l_o^2}}\left[1-\gamma(x,x')O(x,x')\right]^{N-1}\, ,
\end{align}
where $\gamma(x,x')$ is defined in Eq.~(\ref{defgamma}). The momentum distribution follows from Eq.~(\ref{defnk}), and results 
for different values of the statistics parameter are shown in Fig.~\ref{fig1}.

Several features are noteworthy. As in the homogeneous case, the momentum distributions are asymmetric in $k$ for generic statistics and exhibit small oscillations due to the trap. 
Unlike the case of trapped free fermions, however, the number of oscillations does not scale directly with the particle number. The momentum width increases with $N$ at fixed oscillator 
length, similar to the behavior observed for $p$-wave fermions \cite{BEG05}, but in contrast to trapped interacting bosons \cite{AG03}. 
At large momentum, the distributions decay as
$ n(k)\rightarrow C_H/k^2\, ,|k|\rightarrow \infty$ reflecting the short-distance structure of the anyonic wavefunction [Eq.~(\ref{shortbehav})].
Note that in this case the Tan contact  $C_H$ is different from the one for the homogeneous system given by Eq.~(\ref{tanhom}). 
In the thermodynamic limit $l_o\pi^{1/2}\rightarrow\infty$, $N\rightarrow\infty$ with fixed density $n=N/(l_o\pi^{1/2})$
the short-distance correlations in the bulk (trap center) reduce to  Eq.~(\ref{s4:exp}), consistent with the expansion $\mbox{Erf}(z/l_o)\sim 2 z/(l_o\pi^{1/2})$ at small arguments \cite{BEG05}.

\begin{figure*}
\includegraphics[width=1\linewidth]{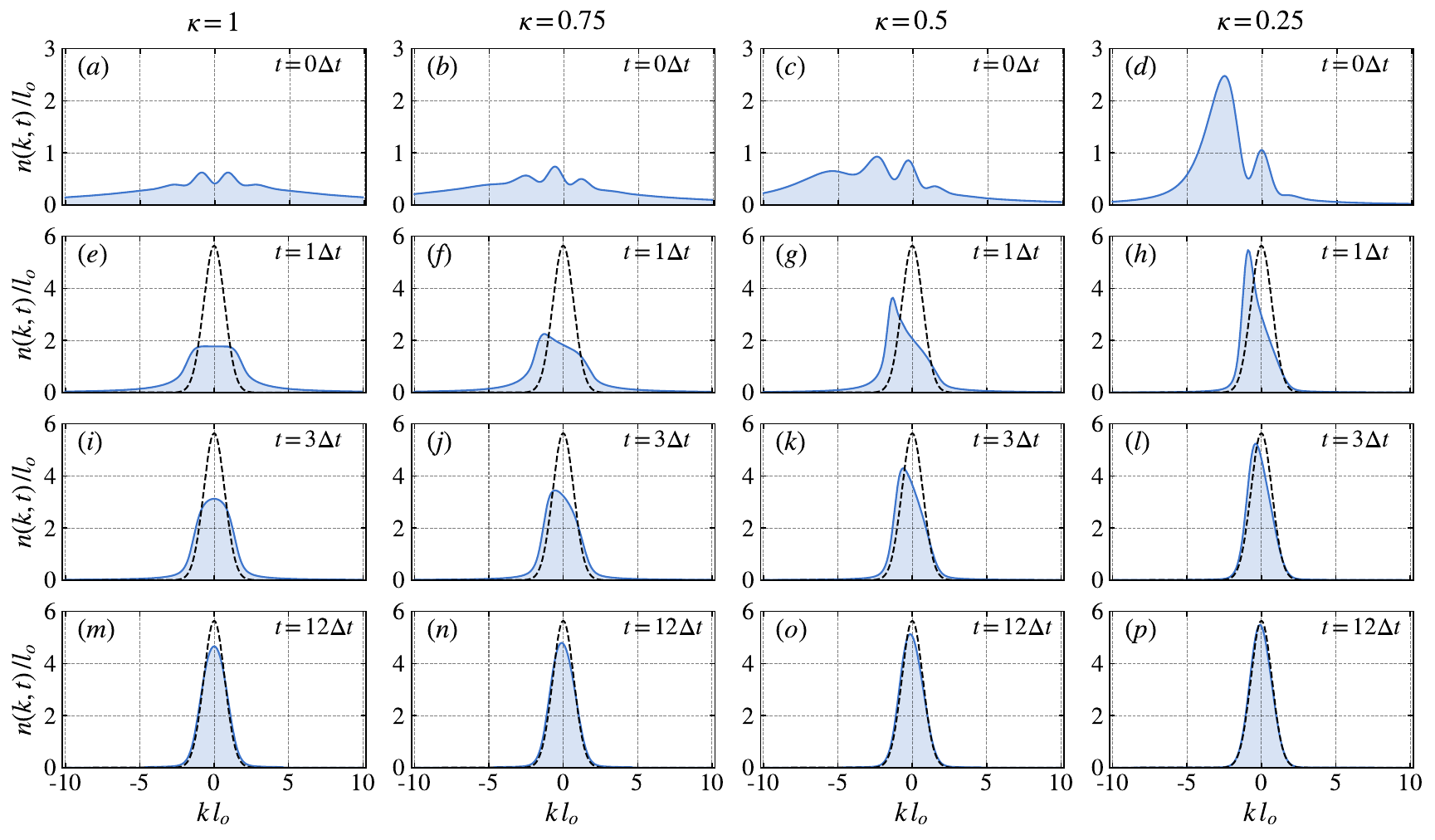}
\caption{
Time evolution of the momentum distribution of $sp$-anyons after release from the trap for various statistics parameters. Parameters are 
$N=10$,  $l_0 = m = \omega_0 = 1$   and $\Delta t=2\pi/\omega_0$. Dashed black lines in the last three rows show the momentum distribution of $N$ free bosons in the initial 
trap [Eq.~(\ref{mombosons})].
}
\label{fig2}
\end{figure*}

\section{Nonequilibrium dynamics}\label{s4}

Ultracold gases provide an invaluable platform for investigating the properties of many-body systems, thanks to the high degree of 
control over interactions, dimensionality, and statistics, both at equilibrium and out of equilibrium. Over the past decades, a wide 
range of nonequilibrium scenarios have been explored in cold-atom setups, revealing exotic and sometimes unexpected phenomena, such 
as the absence of thermalization in integrable and near-integrable systems.  

The time evolution of $sp$-anyons confined in a harmonic trap with time-dependent frequency,  
\be
V(z)=\tfrac{1}{2} m \omega(t)^2 z^2 ,
\ee
can be studied both analytically and numerically by exploiting the fact that the dynamics of Hermite functions is governed by \cite{PP70,PZ98} 
\be
\phi_j(z,t)=\frac{1}{\sqrt{b}}\, \phi_j\!\left(\frac{z}{b},0\right) 
\exp\left[i\frac{mz^2}{2}\frac{\dot b}{b}-i\varepsilon(0)\tau(t)\right],
\ee
where $b(t)$ is the solution of the Ermakov--Pinney equation
\be
\ddot{b}+\omega(t)^2 b=\frac{\omega_0}{b^3},
\ee
with initial conditions $b(0)=1$, $\dot{b}(0)=0$, $\varepsilon(0)=\omega_0(j+1/2)$, and $\omega_0=\omega(t\le 0)$. The rescaled time parameter $\tau(t)$ is defined as
\be
\tau(t)=\int_0^t \frac{dt'}{b(t')^2}.
\ee

In the groundstate, described by Eq.~(\ref{wavegs}), the dynamics of the one-particle 
reduced density matrix  is given by
\be\label{rdmdynamics}
\rho^{(1)}(x,x'|t)=\frac{1}{b}\,\rho^{(1)}\!\left(\frac{x}{b},\frac{x'}{b}\bigg|0\right) 
\exp\left[-\frac{i}{ b} \frac{\dot{b}}{\omega_0}\frac{x^2-(x')^2}{2 l_0^2}\right].
\ee
From this relation, it follows that the evolution of the density profile is particularly simple: it coincides with that of free bosons subjected to the same quench,
\be\label{densdynamics}
\rho^{(1)}(x|t)=\rho^{(1)}_B(x|t)=\frac{1}{b}\,\rho^{(1)}\!\left(\frac{x}{b},\frac{x}{b}\bigg|0\right).
\ee
By contrast, the momentum distribution exhibits nontrivial dynamics,
\begin{align}\label{momdynamics}
n(k,t)=&\, b\int \!\!\int dx\, dx'\,\rho^{(1)}(x,x'|0)  \nonumber \\
&\times \exp\left[-i b \left(\frac{\dot{b}}{\omega_0}\frac{x^2-(x')^2}{2 l_0^2}-i k(x-x')\right)\right].
\end{align}
In the following sections, we analyze two paradigmatic nonequilibrium scenarios:  
i)  the expansion of $sp$-anyons after release from a harmonic trap, and  
ii) the oscillations induced by a sudden change in the trap frequency.

\subsection{Dynamical bosonization}

\begin{figure*}
\includegraphics[width=1\linewidth]{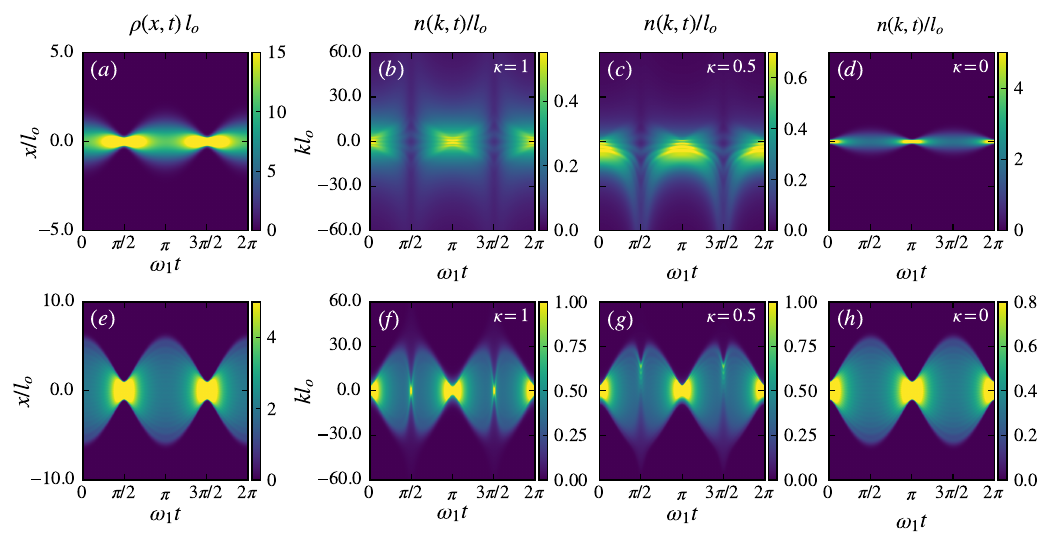}
\caption{
Density profiles (first column) and momentum distribution oscillations (last three columns) induced by a quench of the trap frequency 
($\omega_0=1$, $\omega_1=6\omega_0$, $\alpha \approx -0.972$, $l_0=m=1$) for systems of $N=20$ $sp$-anyons with different statistics 
parameters (top row), and for a comparable system of Tonks-Giradeau anyons described by Eq.~(\ref{swave}) (bottom row).
}
\label{fig3}
\end{figure*}

Free expansion following the release from a trapping potential is one of the most commonly encountered nonequilibrium scenarios in the study of many-body 
ultracold gases. In this section, we show analytically that after release from the harmonic trap: 
i) the asymptotic momentum distribution at long times has the same shape as the density profile of free bosons in the initial trap, and 
ii) the asymptotic momentum distribution coincides with that of free bosons in the initial trap.

This phenomenon is referred to as \emph{dynamical bosonization} and was first discovered in \cite{GM06}, where the free expansion of $p$-wave 
fermions was investigated. It is worth mentioning that in the case of strongly repulsive bosons (Tonks-Girardeau gas), a similar phenomenon 
occurs, in which the asymptotic quantities are related to those of free fermions in the initial trap. This effect, known as 
\emph{dynamical fermionization}, was predicted in \cite{RM05, MG05} (see also \cite{delC08,GP08,BHLM12,CGK15,XR17,CDDK19,ASYP21,Patu23}) 
and experimentally confirmed in \cite{WMLZ20}.  

To proceed, it is useful to first examine the density profile and momentum distribution of free bosons in the initial trap, as these are the 
quantities expected to be related to the asymptotic behavior of $sp$-anyons. The density profile is given by (see Eq.~(\ref{densb})) 
\be\label{s4:e2}
\rho_B^{(1)}(x)=N|\phi_0(x)|^2\,,
\ee
while the momentum distribution is
\be
n_B(k)=\frac{1}{2\pi}\int\int e^{-i k(x-x')} N \phi_0^*(x') \phi_0(x)\, dx\, dx'\,.
\ee
Using the Fourier transform formula $ \frac{1}{\sqrt{2\pi}} \int e^{i k x} \phi_0(x)\, dx = l_o \phi_0(k l_o^2),$
where $\phi_0(x)$ is given by Eq.~(\ref{defphi0}), one obtains
\be\label{mombosons}
n_B(k) = N l_o^2 \, \left| \phi_0(k l_o^2) \right|^2.
\ee
This relation also shows that the momentum distribution of trapped free bosons has the same shape as the density profile $n_B(k)=l_o^2\, \rho_B^{(1)}(k l_o^2).$

We now investigate the asymptotic momentum distribution of $sp$-anyons at long times after release from the trap using the method of stationary 
phase \cite{MG05,GM06}. The solution of 
the Ermakov-Pinney equation for free expansion, $\omega(t)=\Theta(-t)\omega_0$, is
$
b(t)=\sqrt{1+\omega_0^2t^2}, \quad \text{with } \lim_{t\to\infty}b(t)\sim\omega_0 t.
$
A stationary phase analysis of the integrals in Eq.~(\ref{momdynamics}) reveals that the stationary points coincide, 
$k_0 = k \, \omega_0 l_o^2 / \dot{b}$, and therefore
\be
\lim_{t \to \infty} n(k,t) \sim l_o^2 \rho^{(1)}(k l_o^2, k l_o^2 | 0) = l_o^2 \rho_B^{(1)}(k l_o^2) = n_B(k),
\ee
proving both assertions stated at the beginning of this section.  
In Fig.~\ref{fig2}, we show the time evolution of the momentum distribution for $N=10$ $sp$-anyons with different statistics parameters 
after release from a harmonic trap with $l_o = m = \omega_0 = 1$. While the initial momentum distributions are highly asymmetric for 
$\kappa \ne 1$, in all cases one can observe that the asymptotic $n(k)$ approaches that of free bosons in the initial trap, as given 
by Eq.~(\ref{mombosons}).

\subsection{Breathing oscillations}

A sudden change of the trap frequency at $t=0$ induces oscillations in the trapped gas. Such oscillations have been experimentally observed 
in single-component bosons \cite{FCJB14,WMLZ20}. If we denote the initial frequency before the quench by $\omega_0$ and the post-quench 
frequency by $\omega_1$, the solution of the Ermakov--Pinney equation is  
\be
b(t) = \left[1 + \alpha \sin^2(\omega_1 t)\right]^{1/2}, 
\qquad \alpha = \frac{\omega_0^2}{\omega_1^2} - 1.
\ee
In this case, $b(t)$ describes oscillations with period $\pi/\omega_1$, with amplitude varying between $1$ and $\omega_0/\omega_1$.  

The oscillations of the density and momentum distribution for a system of $N=20$ $sp$-anyons are shown in the first row of Fig.~\ref{fig3}, 
for a quench with $\omega_0=1$ and $\omega_1=6\omega_0$. The density, which is independent of statistics and coincides with that of free bosons 
[Eqs.~(\ref{s4:e2}) and (\ref{densdynamics})], reaches its maximum width at $\omega_1 t = p \pi,$ $p=0,1,2,\dots$ and minimum width at 
$\omega_1 t = (p+1/2)\pi,$ $p=0,1,2,\dots$. By contrast, the momentum distribution exhibits the opposite behavior: it has maximum width at 
$\omega_1 t=(p+1/2)\pi$ and minimum width at $\omega_1 t = p\pi$.  

It is instructive to compare these results for $sp$-anyons with those for Tonks-Girardeau anyons, whose wavefunctions are
\be\label{swave}
\psi_s(\bz_N) = A_\kappa(\bz_N)\,\psi_-^{(0)}(\bz_N),\ 
\psi_-^{(0)}(\bz_N)=\frac{1}{\sqrt{N!}} \det_N[\phi_i(z_j)] ,
\ee
where $\psi_-^{(0)}(\bz_N)$ is the Slater determinant constructed from the first $N$ single-particle orbitals. At $\kappa=1$, the Tonks-Girardeau 
anyons describe the bosonic Tonks-Girardeau gas (impenetrable bosons) \cite{Gira60}, while for $\kappa=0$ they reduce to free fermions.  

The dynamics of Tonks-Girardeau anyons are shown in the second row of Fig.~\ref{fig3}, for a comparable particle number and under the same frequency quench. 
The density evolution resembles that of the $p$-wave case, although the $s$-wave density, identical to the free-fermion result, is broader and given by
$\rho_s^{(1)}(x) = \sum_{j=0}^{N-1} |\phi_j(x)|^2 \, .$
In contrast, the momentum distribution exhibits an additional narrowing at times $\omega_1 t = (p+1/2)\pi$, with $p=0,1,2,\dots$. This effect is a genuine 
many-body phenomenon, first identified in Tonks-Girardeau bosons \cite{ABGK17} and later generalized to anyons in \cite{Patu20}. Physically, it can be 
understood as a self-reflection of the cloud arising from strong repulsive interactions.

\section{Universality of the RDMs}\label{s5}

The importance of the reduced density matrices  defined in Eq.~(\ref{defrdm}) cannot be overstated. They serve as essential 
tools in determining density profiles, momentum distributions, entanglement between subsystems, the presence of off-diagonal 
long-range order \cite{Yang62}, collective formation of pairs \cite{Yang62} or triples \cite{GT22a,GT23}, and in identifying 
$n$-order coherence and fragmentation phenomena \cite{SSAC08,Lode16}.

In many physical applications, it is useful to express the $n$-particle RDM in diagonal form (here we omit the dependence on 
the state $\boldsymbol{j}$):
\be
\rho^{(n)}(\bx_n,\bx_n') = \sum_k \lambda_k\, u_k^*(\bx_n') u_k(\bx_n) \, ,
\ee
where the eigenfunctions $u_k(\bx_n)$ and eigenvalues $\lambda_k$ satisfy the integral equation:
\be
\int \rho^{(n)}(\bx_n,\bx_n')\, u_k(\bx_n')\, d\bx_n' = \lambda_k u_k(\bx_n) \, .
\ee
In general, the eigenfunctions are referred to as the natural $n$-functions, and the eigenvalues as the natural occupations. 
In the special case of $n=1$, the eigenfunctions are known as natural orbitals, and in the absence of an external potential, 
the eigenvalues are directly related to the momentum distribution.

For the groundstate described by Eq.~(\ref{wavegs})
the $n$-particle RDM takes the form:
\begin{align}
\rho^{(n)}(\bx_n,\bx_n') &= \frac{N!}{(N-n)!} A_{-\kappa}(\bx_n')\, A_{\kappa}(\bx_n)\, \prod_{j=1}^n \phi_0^*(x_j') \phi_0(x_j) \nonumber \\
&\qquad\qquad\qquad \times \left[P(\bx_n,\bx_n')\right]^{N-n} \, ,
\end{align}
with
\be
P(\bx_n,\bx_n') = \int dz\, \phi_0^*(z) \phi_0(z)\, \prod_{k=1}^n e^{-i\frac{\pi\kappa}{2}\left[\epsilon(z - x_k') - \epsilon(z - x_k)\right]} \, .
\ee
As these expressions depend on $\phi_0(z)$ in a non-trivial way, one would naturally expect the RDMs to be strongly influenced 
by the external confinement potential. However, in a remarkable paper, Ko\'{s}cik and Sowi\'{n}ski \cite{KS23} demonstrated that 
for $p$-wave fermions and for any value of $n$, the eigenvalues of the RDMs are independent of the confinement potential. 
Furthermore, the corresponding eigenfunctions under different confinements are related via a simple analytical transformation.
In this section, we prove that a similar universal behavior emerges in the RDMs of $sp$-anyons.

\subsection{Universal form of the one-particle RDM}

We begin by determining the universal form of the one-particle reduced density matrix, following the method introduced in \cite{KS23}. 
As we shall see, this case already contains the core ideas and techniques necessary to address the general $n$-particle case.

In the groundstate, the one-particle RDM is given by
\begin{align}\label{1rdmgs}
\rho^{(1)}(x,x') = N \phi_0^*(x') \phi_0(x) \left[P(x,x')\right]^{N-1} \, ,
\end{align}
where
\be
P(x,x') = \int dz\, \phi_0^*(z) \phi_0(z) e^{-i\frac{\pi\kappa}{2} \left[\epsilon(z - x') - \epsilon(z - x)\right]} \, .
\ee
The integral equation for the natural orbitals is
\be\label{1eigen}
\int \rho^{(1)}(x,x')\,u_k(x')\, dx'=\lambda_k u(x)\, .
\ee
Throughout this section, unless explicitly stated otherwise, all integrals are taken over the interval $[L_-, L_+]$.

We now define the partial overlap function
\be
F(x) = \int_{L_-}^x \phi_0^*(z) \phi_0(z)\, dz\, ,
\ee
which is strictly increasing, with $F(L_-)=0$ and $F(L_+)=1$. Hence, $F(x)$ is a bijective function with inverse $F^{-1}(x)$.

By changing variables in Eq.~(\ref{1rdmgs}) via $\xi = F(z)$, with $d\xi = \phi_0^*(z)\phi_0(z)dz$, the integral equation \eqref{1eigen} becomes:
\begin{widetext}
\begin{align}\label{s5:e1}
N\int\phi_0^*(x')\phi_0(x) \left(\int_0^1 d\xi\, e^{-i\frac{\pi\kappa}{2}\left[\epsilon\left(F^{-1}(\xi)-x'\right)-\epsilon\left(F^{-1}(\xi)-x\right)\right]}\right)^{N-1}
u_k(x')\, dx'=\lambda_k u_k(x)\, .
\end{align}
\end{widetext}
We now introduce a transformation $v_k(x)$ defined by
\be\label{vectort}
u_k(x) = \phi_0(x)\, v_k[F(x)]\, ,
\ee
which is equivalent to $u_k[F^{-1}(x)]/\phi_0[F^{-1}(x)]=v_k(x)\, .$
This transformation is unitary, preserving the norm of the natural orbitals:
\begin{align}
\int u_k(x)^* u_{k'}(x)\, dx&=\int |\phi_0^*(x)|^2 \left(v_{k}[F(x)]\right)^*v_{k'}[F(x)]\, dx\, ,\nonumber\\
&=\int_0^1 v_k^*(y)v_{k'}(y)\, dy\, .
\end{align}
Applying the change of variables $y' = F(x')$ (so that $dy'=\phi_0^*(x')\phi_0(x')dx'$), and setting $y = F(x)$ in Eq.~(\ref{s5:e1}), 
the eigenvalue equation becomes
\be\label{universal1rdmp}
\int_0^1 \rho_0^{(1)}(y, y') v_k(y')\, dy' = \lambda_k v_k(y)\, ,
\ee
where $\rho_0^{(1)}(y,y')$ is the universal one-particle RDM:
\begin{widetext}
\begin{align}
\rho_0^{(1)}(y,y')&= N\left(\int_0^1 d\xi\, e^{-i\frac{\pi\kappa}{2}\left[\epsilon\left(F^{-1}(\xi)-F^{-1}(y')\right)-\epsilon\left(F^{-1}(\xi)-F^{-1}(y)\right)\right]}\right)^{N-1}\, .
\end{align}
\end{widetext}
Using the identity $\epsilon[F^{-1}(a)-F^{-1}(b)] =\epsilon(a - b)$, which follows from the strict monotonicity of $F$, we can simplify the expression to:
\begin{align}\label{universal1rdm}
\rho_0^{(1)}(y,y')= N\left(1-\left[1-e^{-i\pi\kappa\epsilon(y-y')}\right]|y-y'|\right)^{N-1}\, .
\end{align}
A key feature of this computation is that the eigenvalues of the original RDM $\rho^{(1)}(x, x')$ are identical to those 
of the universal RDM $\rho_0^{(1)}(y, y')$, which is independent of the external potential through $\phi_0(z)$. This 
significantly simplifies the task of determining the eigenvalues and natural orbitals. Once the eigenfunctions $v_k(y)$ 
of the universal eigenproblem \eqref{universal1rdmp} are obtained, the corresponding eigenfunctions of the original problem 
are readily constructed using the transformation (\ref{vectort}).

\subsection{Universal form of the $n$-particle RDM}

Deriving the universal form of the $n$-particle RDM follows the same pattern as in the one-particle case. 
We now perform simultaneously the change of variables
\be
y_j=F(x_j)\, ,\ \ y_j'=F(x_j')\, ,\ \  j=1,\cdots,n\, ,
\ee
together with the transformation of the eigenfunctions
\be\label{s5:transf}
u_k(\bx_n)=\left(\prod_{j=1}^n\phi_0(x_j)\right)v_k\left[F(x_1),\cdots,F(x_n)\right]\, .
\ee
Using the strict monotonicity of $F$, the eigenvalue problem for the $n$-particle RDM reduces to the universal form
\be\label{s5:e2}
\int_0^1\rho_0^{(n)}(\boldsymbol{y}_n,\boldsymbol{y}_n')\, v_k(\boldsymbol{y}_n')\, d\boldsymbol{y}_n'=\lambda_k v_k(\boldsymbol{y}_n)\, ,
\ee
with
\be\label{s5:e3}
\rho_0^{(n)}(\boldsymbol{y}_n,\boldsymbol{y}_n')=\frac{N!}{(N-n)!}A_{-\kappa}(\boldsymbol{y}_n')A_{\kappa}(\boldsymbol{y}_n) \left[P_0(\boldsymbol{y}_n,\boldsymbol{y}_n')\right]^{N-n}\, ,
\ee
where
\be
P_0(\boldsymbol{y}_n,\boldsymbol{y}_n')=\int_0^1 d\xi\, \prod_{j=1}^ne^{-i\frac{\pi\kappa}{2}\left[\epsilon\left(\xi-y_j'\right)-\epsilon\left(\xi-y_j\right)\right]}\, .
\ee
Equations~(\ref{s5:e2}) and (\ref{s5:e3}) provide the universal formulation of the $n$-particle RDM eigenproblem. 
Crucially, the eigenvalues are independent of the trapping potential, and the corresponding eigenfunctions in the original 
coordinates are recovered via the transformation (\ref{s5:transf}).

\section{Solution of the eigenproblem for the one-particle RDM}\label{s6}

Even for relatively simple systems, solving the eigenproblem (\ref{1eigen}) and analytically determining the natural orbitals together with their associated 
eigenvalues (natural occupations) is typically a challenging task. Nevertheless, the polynomial structure of the universal one-particle RDM 
(\ref{universal1rdm}) indicates that a complete analytical solution may be within reach.

For the case of $p$-wave fermions ($\kappa=1$) with $N=2$ particles, the natural orbitals and occupation numbers were determined in \cite{KS23},
where it was found that the eigenvalues display a twofold degeneracy. Subsequently, Sabater {\it et al.} \cite{SRAJ24} demonstrated numerically that the
eigenvectors identified in \cite{KS23} remain valid for larger $N$, with the corresponding eigenvalues again appearing in degenerate pairs. Moreover, they
 derived explicit analytic expressions for the eigenvalues at arbitrary $N$ using a recurrence relation.

In this section, we present a direct solution of the eigenproblem (\ref{universal1rdmp}), obtaining closed-form expressions for both the eigenvalues and 
the natural orbitals for arbitrary values of the statistics parameter. The $p$-wave fermion case analyzed in \cite{KS23,SRAJ24} then appears as a special 
instance of our general solution, which also reveals the origin of the double degeneracy and clarifies the specific structure of the natural orbitals.

\subsection{Boundary conditions}

Our strategy for solving the eigenproblem consists in deriving an ordinary differential equation (ODE) with constant coefficients for the 
natural orbitals, starting from Eq.~(\ref{universal1rdmp}). While the general solution of such ODEs is standard, the essential step is to 
determine the boundary conditions satisfied by the natural orbitals and their derivatives.

Adopting slightly modified notations, the universal eigenproblem for the one-particle RDM (\ref{universal1rdmp}) can be recast in the form 
\begin{align}\label{eigeneq}
\int_0^x & N\left(1- a x+a y\right)^{N-1} v(y)\, dy\nonumber\\
&\ \ +\int_x^1 N\left(1+\overline{a} x- \overline{a} y\right)^{N-1} v(y)\, dy=\lambda\, v(x)\, ,
\end{align}
where
\be
a=1-e^{- i\pi\kappa}\, ,\ \ \ \overline{a}=a=1-e^{ i\pi\kappa}\, .
\ee
Evaluating (\ref{eigeneq}) at $x=0$ and $x=1$ yields
\begin{align}
\int_0^1 N\left(1-\overline{a}y\right)^{N-1}v(y)\, dy&=\lambda\, v(0)\, ,\\
\int_0^1 N\left(1-a +ay\right)^{N-1} v(y)\, dy&=\lambda\, v(1)\, .
\end{align}
Since $1-a+ay=e^{-i\pi\kappa}(1-\overline{a} y)$, the above relations imply the boundary condition
\be\label{bceigen}
v(0)e^{-i\pi\kappa(N-1)}=v(1)\, .
\ee
In Appendix~\ref{app:odebc} we further establish that analogous boundary conditions hold for all derivatives of the natural orbitals. 
Specifically, if $v^{(n)}(x)$ denotes the $n$-th derivative of $v(x)$, then
\be\label{bcgeneral}
v^{(n)}(0)e^{-i\pi\kappa(N-1)}=v^{(n)}(1)\, ,\ \  n=1,\cdots,N-1\, .
\ee

\subsection{ODE for the natural orbitals}

By taking successive derivatives of Eq.~(\ref{eigeneq}), as detailed in Appendix~\ref{app:odebc}, one finds that the eigenvectors of the 
one-particle RDM satisfy the following ordinary differential equation:
\begin{align}
\sum_{j=0}^{N-1}\frac{N!}{(N-j-1)!}&\tilde{C}(\kappa,j)\,v^{(N-j-1)}(x)=\lambda\, v^{(N)}(x)\, ,\label{ode}
\end{align}
with 
\be
\tilde{C}(\kappa,j)=(-2 i)^{j+1}\left[\sin\left(\frac{\pi\kappa}{2}\right)\right]^{j}\sin\left(\frac{\pi\kappa j}{2}\right)\, .
\ee
Equation~(\ref{ode}) is an $N$-th order linear ODE with constant coefficients. Denoting by $\alpha_1, \dots, \alpha_N$ the roots of the associated 
characteristic equation
\be\label{chareq}
\sum_{j=0}^{N-1}\frac{N!}{(N-j-1)!}\tilde{C}(\kappa,j)\,\alpha^{N-j-1}=\lambda\, \alpha^N\, ,
\ee
the general solution of the ODE is of the form  $\sum_{j=1}^N c_j e^{\alpha_j x}$ if all $\alpha_j$ are distinct and more 
generally  $\sum_{j=1}^N \left(\sum_{l=1}^{k_j} c_{j,l}x^{l-1}\right)e^{\alpha_j x}$ if 
$\alpha_j$ is a root of multiplicity  $k_j$ where $c_j$ and $c_{j,l}$ are constants. In Appendix~\ref{app:absence} it is shown that 
the characteristic equation admits no repeated roots and that all solutions are purely imaginary. Consequently, the eigenvectors take 
the form 
\be\label{norbitals}
v_k(x)=e^{i \alpha_k x}\, ,\ \   \alpha_k=2\pi k-\delta\, ,\  \ k=0,\pm1\cdots\, ,
\ee
where $\delta$ is determined by
\be
\pi\kappa(N-1)=2\pi r +\delta\, ,
\ee
with $r$ the nearest integer. In particular, for $p$-wave fermions ($\kappa=1$) one finds $\delta=\pi$ when $N$ is even and $\delta=0$ when $N$ is odd.
The corresponding eigenvalues $\lambda_k$ are given by
\begin{align}\label{eigenvalues}
\sum_{j=0}^{N-1}\frac{N!}{(N-j-1)!}\,\frac{C(\kappa,j)}{(2\pi k-\delta)^{j+1}}=\lambda_k\, ,
\end{align}
with 
\be\label{defc}
C(\kappa,j)=(-2 )^{j+1}\left[\sin\left(\frac{\pi\kappa}{2}\right)\right]^{j}\sin\left(\frac{\pi\kappa j}{2}\right)\, .
\ee
We have thus obtained an explicit analytical solution of the eigenproblem for the universal one-particle RDM. The natural orbitals are given by 
(\ref{norbitals}), with the associated eigenvalues specified in (\ref{eigenvalues}). In the following, we examine particular cases and establish 
connections with the results discussed at the beginning of this section.

\subsection{$p$-wave fermions ($\kappa=1$)}

Results for $p$-wave fermions are recovered for $\kappa=1$. In this case, from Eq.~(\ref{defc}) we have 
$C(\kappa,j)=(-2)^{j+1}\sin\left(\frac{\pi j}{2}\right)$,
and one must distinguish two cases depending on the parity of $N$.

{\it N even.}  When $N$ is even, $\delta=\pi$ and the eigenvalues are doubly degenerate, given by 
\be\label{eigenneven}
\lambda_k=\sum_{m=1}^{N/2}\frac{N!}{(N-2m)!}\frac{(-1)^{m+1}4^m}{\left[(2k-1)\pi\right]^{2m}}\, ,\ k=0,1,\cdots\, ,
\ee
with eigenvectors
\be
v_{k\pm}(x)=e^{\pm i (2k-1)\pi x}\, ,\ \ k=0,1,\cdots\, .
\ee
For each $k$, one may can take linear combinations of $v_{k\pm}(x)$ yielding
\begin{subequations}\label{s5:e5}
\begin{align}
\tilde{v}_{k+}(x)&=\sqrt{2}\sin[(2k-1)\pi x]\, ,\\
\tilde{v}_{k-}(x)&=\sqrt{2}\cos[(2k-1)\pi x]\, .
\end{align}
\end{subequations}
The eigenvalues (\ref{eigenneven}) and natural orbitals (\ref{s5:e5}) coincide with the results presented in \cite{KS23,SRAJ24} for even $N$.

{\it N odd.} When $N$ is odd, $\delta=0$. The case $k=0$ is special, with $v_0(x)=1$ and eigenvalue $\lambda_0=1$. For $k\ne 0$ there is again double degeneracy with 
eigenvalues 
\be\label{eigennodd}
\lambda_k=\sum_{m=1}^{(N-1)/2}\frac{N!}{(N-2m)!}\frac{(-1)^{m+1}4^m}{\left[2 k\pi\right]^{2m}}\, ,\ k=1,2,\cdots\, ,
\ee
and natural orbitals
\be
v_{k\pm}(x)=e^{\pm i 2k\pi x}\, ,\ \ k=1,2,\cdots\, .
\ee
Linear combinations yield
\begin{subequations}\label{s5:e6}
\begin{align}
\tilde{v}_{k+}(x)&=\sqrt{2}\sin[2k\pi x]\, ,\\
\tilde{v}_{k-}(x)&=\sqrt{2}\cos[2k\pi x]\, .
\end{align}
\end{subequations}
in agreement with the results reported in \cite{SRAJ24}.

\subsection{Free bosons ($\kappa=0$)}

For $\kappa=0$, we have $C(\kappa,j)=0$ for all $j$. The eigenvalue problem reduces to
$\int_0^1 Nv(y)\, dy=\lambda v(x)$ with  natural orbital $v(x)=1$ and eigenvalue $\lambda=N$, 
consistent with the expected groundstate of a system of free bosons.

\section{Conclusions}

We have investigated a system of strongly interacting one-dimensional anyons, representing the generalization of spin-polarized fermions 
with $p$-wave interactions to arbitrary statistics. This system exhibits an asymmetric momentum distribution, which in the homogeneous 
case takes the form of a shifted Lorentzian. Following release from a harmonic trap, $sp$-anyons undergo dynamical bosonization, 
while a sudden quench of the trap frequency induces breathing oscillations without the additional narrowing of the momentum distribution 
observed for Tonks-Girardeau anyons. We further proved that the $n$-particle RDMs are universal, independent 
of the external confinement, and we analytically solved the eigenproblem for the one-particle RDM. A natural extension of this work is 
to explore regimes beyond strong interactions, where the wavefunction can be constructed from the Lieb-Liniger Bethe ansatz eigenstates
combined with  the anyonic Bose-Fermi mapping. This direction will be pursued in future work.

\vspace{1cm}
\acknowledgments

O.I.P. acknowledges financial support from Grant No. 30N/2023, provided through the National Core Program of the 
Romanian Ministry of Research, Innovation, and Digitization.

\appendix

\begin{widetext}

\section{Derivation of the ODE and boundary conditions for the natural orbitals}\label{app:odebc}

Starting from Eq.~(\ref{eigeneq}) successive applications of the Leibniz integral rule
\begin{align}
\frac{d}{dx}\left(\int_{a(x)}^{b(x)}f(x,y)\, dy\right)=f[x,b(x)]\frac{d b(x)}{dx}-f[x,a(x)]\frac{da(x)}{dx}+\int_{a(x)}^{b(x)}\frac{\partial f(x,y)}{\partial x}\, dy\, ,
\end{align}
produce the following expressions for the first three derivatives of the eigenvectors:
\begin{align}\label{firstder}
&N\left[(-a)^0-(\overline{a})^0\right]v(x)\nonumber\\
&\ \ +\frac{N!}{(N-2)!}\left[(-a)\int_0^x(1-ax+ay)^{N-2} v(y)\, dy+(\overline{a})\int_x^1(1+\overline{a} x-\overline{a} y)^{N-2}v(y)\, dy \right] =\lambda\, v^{(1)}(x)\, ,
\end{align}
\begin{align}\label{secondder}
&N\left[(-a)^0-(\overline{a})^0\right]v^{(1)}(x)+\frac{N!}{(N-2)!}\left[(-a)^1-(\overline{a})^1\right]v(x)\nonumber\\
&+\frac{N!}{(N-3)!}\left[(-a)^2\int_0^x(1-ax+ay)^{N-3} v(y)\, dy+(\overline{a})^2\int_x^1(1+\overline{a} x-\overline{a} y)^{N-3}v(y)\, dy \right]=\lambda\, v^{(2)}(x)\, ,
\end{align}
\begin{align}
&N\left[(-a)^0-(\overline{a})^0\right]v^{(2)}(x)+\frac{N!}{(N-2)!}\left[(-a)^1-(\overline{a})^1\right]v^{(1)}(x)+\frac{N!}{(N-3)!}\left[(-a)^2-(\overline{a})^2\right]v(x)\nonumber\\
&+\frac{N!}{(N-4)!}\left[(-a)^3\int_0^x(1-ax+ay)^{N-4} v(y)\, dy+(\overline{a})^3\int_x^1(1+\overline{a} x-\overline{a} y)^{N-4}v(y)\, dy \right]=\lambda\, v^{(3)}(x)\, .
\end{align}
The emerging pattern reveals that the eigenvectors satisfy the $N$-th order ODE
\begin{align}
\sum_{j=0}^{N-1}\frac{N!}{(N-j-1)!}&\left[(-a)^j-(\overline{a})^j\right]v^{(N-j-1)}(x)=\lambda\, v^{(N)}(x)\, ,
\end{align}
which is equivalent to Eq.~(\ref{ode}).

Now let us look at the boundary conditions. 
The natural orbitals satisfy Eq.~(\ref{bceigen}). Evaluating the first derivative (\ref{firstder}) at $x=0$ and $x=1$
gives
\begin{align}
N\left[(-a)^0-(\overline{a})^0\right]v(0)+\frac{N!}{(N-2)!}(\overline{a})\int_0^1(1-\overline{a}y)^{N-2}v(y)\, dy=\lambda\, v^{(1)}(0)\, ,\\
N\left[(-a)^0-(\overline{a})^0\right]v(1)+\frac{N!}{(N-2)!}(-a)\int_0^1(1-a+ay)^{N-2}v(y)\, dy=\lambda\, v^{(1)}(1)\, .
\end{align}
Using $1-a+ay=e^{-i\pi\kappa}(1-\overline{a} y)$  in the second equation, we can write
\begin{align} 
N\left[(-a)^0-(\overline{a})^0\right]v(1)+\frac{N!}{(N-2)!}(-a)e^{-i\pi\kappa (N-2)}\int_0^1(1-\overline{a}y)^{N-2}v(y)\, dy=\lambda\, v^{(1)}(1)\, , 
\end{align}
while from the first equation we have $\int_0^1(1-\overline{a}y)^{N-2}v(y)\, dy=\frac{(N-2)!}{N!}\left(\lambda\, v^{(1)}(0)-N\left[(-a)^0-(\overline{a})^0\right]v(0)\right)$.
Combining the above with  $-a/\overline{a}=e^{-i\pi\kappa}$ and Eq.~(\ref{bceigen}) leads to the first-derivative boundary condition:
\be\label{bcfirst}
v^{(1)}(0)e^{-i\pi\kappa(N-1)}=v^{(1)}(1)\, .
\ee

For the second derivative, evaluating Eq.~(\ref{secondder}) at $x=0$ and $x=1$ gives
\begin{align}
&N\left[(-a)^0-(\overline{a})^0\right]v^{(1)}(0)+\frac{N!}{(N-2)!}\left[(-a)^1-(\overline{a})^1\right]v(0)
+\frac{N!}{(N-3)!}\left[(\overline{a})^2\int_0^1(1-\overline{a} y)^{N-3}v(y)\, dy \right]=\lambda\, v^{(2)}(0)\, ,\\
&N\left[(-a)^0-(\overline{a})^0\right]v^{(1)}(1)+\frac{N!}{(N-2)!}\left[(-a)^1-(\overline{a})^1\right]v(1)
+\frac{N!}{(N-3)!}\left[(-a)^2\int_0^1(1-a+a y)^{N-3}v(y)\, dy \right]=\lambda\, v^{(2)}(1)\, .
\end{align}
Using $\int_0^1(1-a+a y)^{N-3}v(y)\, dy =e^{-i\pi\kappa (N-3)}\int_0^1(1-\overline{a} y)^{N-3}v(y)\, dy$ and  $(-a/\overline{a})^2=e^{-i 2\pi\kappa}$ together with 
Eqs.~(\ref{bceigen}) and (\ref{bcfirst}), yields the second-derivative boundary condition:
\be\label{bcsecond}
v^{(2)}(0)e^{-i\pi\kappa(N-1)}=v^{(2)}(1)\, .
\ee
By induction, the $n$-th derivative satisfies the boundary condition:
\begin{align}\label{bcgeneral_app}
v^{(n)}(0) e^{-i \pi \kappa (N-1)} = v^{(n)}(1),
\end{align}
which corresponds to Eq.~(\ref{bcgeneral}).

\end{widetext}

\section{Absence of repeated solutions}\label{app:absence}

The general solution of Eq.~(\ref{ode}), the ODE satisfied by the natural orbitals of the universal one-particle RDM, can be written as
$\sum_{j=1}^N c_j e^{\alpha_j x}$ if all $\alpha_j$ are distinct solutions of the characteristic equation (\ref{chareq}), and more generally as
$\sum_{j=1}^N \left(\sum_{l=1}^{k_j} c_{j,l}x^{l-1}\right)e^{\alpha_j x}$ when a solution $\alpha_j$ has multiplicity $k_j>1$ where $c_j$ and $c_{j,l}$ are constants.
In this appendix, we show that, due to the boundary conditions (\ref{bcgeneral}): i) all $\alpha_j$ are purely imaginary and ii) no solution has multiplicity  $k_j>1$.
The most general situation occurs when $k_j=N$, in which case the associated eigenvector takes the form
\be
v(x)\equiv\left(c_0+c_1 x+\cdots+c_{N-1}x^{N-1}\right)e^{\alpha x}=p(x)e^{\alpha x}\, .
\ee
The $n$-th derivative of $v(x)$ can be obtained using the general Leibniz rule with the result
\be
v^{(n)}(x)=\sum_{k=0}^nC^n_k p^{(n-k)}(x)\alpha^k e^{\alpha x}\, .
\ee
By induction, the boundary conditions (\ref{bceigen}) and (\ref{bcgeneral}) imply that
\be\label{app:sys}
e^{-i\pi\kappa(N-1)}p^{(n)}(0)=p^{(n)}(1)e^\alpha\, , \ n=0,\cdots,N-1\, .
\ee
Equations~(\ref{app:sys}) imply an upper-triangular system of linear equations for the coefficients $c_0,\cdots,c_{N-1}$.
From this system, one concludes that: i) $e^{\alpha}=e^{-i\pi\kappa(N-1)}$ and ii) all coefficients $c_0,\cdots, c_{N-1}$ must vanish leaving only $c_0$ nonzero.
To illustrate, consider the case $N=4$, for which $p(x)=c_0+c_1 x+c_2 x^2+c_3 x^3$. Then, (\ref{app:sys}) yields
\begin{align}
e^{-i3\pi\kappa}c_0&=(c_0+c_1+c_2+c_3)e^{\alpha}\, ,\\
e^{-i3\pi\kappa}c_1&=(c_1+2c_2+3c_3)e^{\alpha}\, ,\\
2e^{-i3\pi\kappa}c_2&=(2c_2+6c_3)e^{\alpha}\, ,\\
6e^{-i3\pi\kappa}c_3&=6c_3e^{\alpha}\, .
\end{align}
This homogeneous system admits a nontrivial solution only if $e^\alpha=e^{-i 3\pi\kappa}$. Substituting this into the first three equations immediately gives 
$c_3=c_2=c_1=0$ leaving $p(x)=c_0$.
The general case follows by the same reasoning: the boundary conditions enforce that all higher-order coefficients vanish, so that repeated roots ($k_j>1$)
cannot occur. Consequently, all solutions of the characteristic equation are simple and purely imaginary.

\bibliography{PWaveAnyons_V3.bib}

@article{GW00,
  title = {{Dark Solitons in a One-Dimensional Condensate of Hard Core Bosons}},
  author = {Girardeau, M. D. and Wright, E. M.},
  journal = {Phys. Rev. Lett.},
  volume = {84},
  issue = {25},
  pages = {5691--5694},
  numpages = {0},
  year = {2000},
  month = {Jun},
  publisher = {American Physical Society},
  doi = {10.1103/PhysRevLett.84.5691},
  url = {https://link.aps.org/doi/10.1103/PhysRevLett.84.5691}
}

@article{Tan08b,
title = {{Large momentum part of a strongly correlated Fermi gas}},
journal = {Annals of Physics},
volume = {323},
number = {12},
pages = {2971-2986},
year = {2008},
issn = {0003-4916},
doi = {https://doi.org/10.1016/j.aop.2008.03.005},
url = {https://www.sciencedirect.com/science/article/pii/S0003491608000432},
author = {Shina Tan}
}

@article{BZ11,
title = {{Tan relations in one dimension}},
journal = {Annals of Physics},
volume = {326},
number = {10},
pages = {2544-2565},
year = {2011},
issn = {0003-4916},
doi = {https://doi.org/10.1016/j.aop.2011.05.010},
url = {https://www.sciencedirect.com/science/article/pii/S0003491611001084},
author = {Marcus Barth and Wilhelm Zwerger}
}

@article{PK17,
  title = {{Universal Tan relations for quantum gases in one dimension}},
  author = {P\^{a}\c{t}u, Ovidiu I. and Kl\"umper, Andreas},
  journal = {Phys. Rev. A},
  volume = {96},
  issue = {6},
  pages = {063612},
  numpages = {10},
  year = {2017},
  month = {Dec},
  publisher = {American Physical Society},
  doi = {10.1103/PhysRevA.96.063612},
  url = {https://link.aps.org/doi/10.1103/PhysRevA.96.063612}
}

@article{OD03,
  title = {{Short-Distance Correlation Properties of the Lieb-Liniger System and Momentum Distributions of Trapped One-Dimensional Atomic Gases}},
  author = {Olshanii, Maxim and Dunjko, Vanja},
  journal = {Phys. Rev. Lett.},
  volume = {91},
  issue = {9},
  pages = {090401},
  numpages = {4},
  year = {2003},
  month = {Aug},
  publisher = {American Physical Society},
  doi = {10.1103/PhysRevLett.91.090401},
  url = {https://link.aps.org/doi/10.1103/PhysRevLett.91.090401}
}

@article{LM77,
author={Leinaas, J. M.
and Myrheim, J.},
title={On the theory of identical particles},
journal={Il Nuovo Cimento B (1971-1996)},
year={1977},
month={Jan},
day={01},
volume={37},
number={1},
pages={1-23},
doi={10.1007/BF02727953},
url={https://doi.org/10.1007/BF02727953}
}

@article{Wilc82,
  title = {{Quantum Mechanics of Fractional-Spin Particles}},
  author = {Wilczek, Frank},
  journal = {Phys. Rev. Lett.},
  volume = {49},
  issue = {14},
  pages = {957--959},
  numpages = {0},
  year = {1982},
  month = {Oct},
  publisher = {American Physical Society},
  doi = {10.1103/PhysRevLett.49.957},
  url = {https://link.aps.org/doi/10.1103/PhysRevLett.49.957}
}

@article{VVGD25,
  author = {Botao Wang and Amit Vashisht and Yanliang Guo and Sudipta Dhar and Manuele Landini and Hanns-Christoph N\"{a}gerl and Nathan Goldman},
  title  = {{Anyonization of bosons in one dimension: an effective swap model}},
  journal = {arXiv:2504.21208},
  year = {2025},
  url = {https://arxiv.org/abs/2504.21208}
}

@article{Kund99,
  title = {{Exact Solution of Double $\delta$ Function Bose Gas through an Interacting Anyon Gas}},
  author = {Kundu, Anjan},
  journal = {Phys. Rev. Lett.},
  volume = {83},
  issue = {7},
  pages = {1275--1278},
  numpages = {0},
  year = {1999},
  month = {Aug},
  publisher = {American Physical Society},
  doi = {10.1103/PhysRevLett.83.1275},
  url = {https://link.aps.org/doi/10.1103/PhysRevLett.83.1275}
}

@article{BGO06,
  title = {{One-Dimensional Interacting Anyon Gas: Low-Energy Properties and Haldane Exclusion Statistics}},
  author = {Batchelor, M. T. and Guan, X.-W. and Oelkers, N.},
  journal = {Phys. Rev. Lett.},
  volume = {96},
  issue = {21},
  pages = {210402},
  numpages = {4},
  year = {2006},
  month = {Jun},
  publisher = {American Physical Society},
  doi = {10.1103/PhysRevLett.96.210402},
  url = {https://link.aps.org/doi/10.1103/PhysRevLett.96.210402}
}

@article{Gira06,
  title = {{Anyon-Fermion Mapping and Applications to Ultracold Gases in Tight Waveguides}},
  author = {Girardeau, M. D.},
  journal = {Phys. Rev. Lett.},
  volume = {97},
  issue = {10},
  pages = {100402},
  numpages = {4},
  year = {2006},
  month = {Sep},
  publisher = {American Physical Society},
  doi = {10.1103/PhysRevLett.97.100402},
  url = {https://link.aps.org/doi/10.1103/PhysRevLett.97.100402}
}

@article{SSC07,
doi = {10.1088/1742-5468/2007/05/L05003},
url = {https://dx.doi.org/10.1088/1742-5468/2007/05/L05003},
year = {2007},
month = {may},
publisher = {},
volume = {2007},
number = {05},
pages = {L05003},
author = {Santachiara, Raoul and Stauffer, Franck and Cabra, Daniel C},
title = {Entanglement properties and momentum distributions of hard-core anyons on a
ring},
journal = {Journal of Statistical Mechanics: Theory and Experiment}
}

@article{CM07,
  author    = {Pasquale Calabrese and Mario Mintchev},
  title     = {{Correlation functions of one-dimensional anyonic fluids}},
  journal   = {Physical Review B},
  volume    = {75},
  pages     = {233104},
  year      = {2007},
  doi       = {10.1103/PhysRevB.75.233104}
}

@article{PKA07,
doi = {10.1088/1751-8113/40/50/004},
url = {https://dx.doi.org/10.1088/1751-8113/40/50/004},
year = {2007},
month = {nov},
publisher = {},
volume = {40},
number = {50},
pages = {14963},
author = {P\^{a}\c{t}u, Ovidiu I and Korepin, Vladimir E and Averin, Dmitri V},
title = {{Correlation functions of one-dimensional Lieb–Liniger anyons}},
journal = {Journal of Physics A: Mathematical and Theoretical}
}

@article{PKA08a,
doi = {10.1088/1751-8113/41/14/145006},
url = {https://dx.doi.org/10.1088/1751-8113/41/14/145006},
year = {2008},
month = {mar},
publisher = {},
volume = {41},
number = {14},
pages = {145006},
author = {P\^{a}\c{t}u, Ovidiu I and Korepin, Vladimir E and Averin, Dmitri V},
title = {{One-dimensional impenetrable anyons in thermal equilibrium: I. Anyonic generalization of Lenard's formula}},
journal = {Journal of Physics A: Mathematical and Theoretical},
}

@article{SC08,
doi = {10.1088/1742-5468/2008/06/P06005},
url = {https://dx.doi.org/10.1088/1742-5468/2008/06/P06005},
year = {2008},
month = {jun},
publisher = {},
volume = {2008},
number = {06},
pages = {P06005},
author = {Santachiara, Raoul and Calabrese, Pasquale},
title = {{One-particle density matrix and momentum distribution function of one-dimensional anyon gases}},
journal = {Journal of Statistical Mechanics: Theory and Experiment}
}

@article{delC08,
  title = {{Fermionization and bosonization of expanding one-dimensional anyonic fluids}},
  author = {del Campo, A.},
  journal = {Phys. Rev. A},
  volume = {78},
  issue = {4},
  pages = {045602},
  numpages = {4},
  year = {2008},
  month = {Oct},
  publisher = {American Physical Society},
  doi = {10.1103/PhysRevA.78.045602},
  url = {https://link.aps.org/doi/10.1103/PhysRevA.78.045602}
}

@article{BFGL08,
doi = {10.1088/1751-8113/41/46/465201},
url = {https://dx.doi.org/10.1088/1751-8113/41/46/465201},
year = {2008},
month = {oct},
publisher = {},
volume = {41},
number = {46},
pages = {465201},
author = {Batchelor, M T and Foerster, A and Guan, X-W and Links, J and Zhou, H-Q},
title = {{The quantum inverse scattering method with anyonic grading}},
journal = {Journal of Physics A: Mathematical and Theoretical}
}

@article{BGK08,
doi = {10.1088/1751-8113/41/35/352002},
url = {https://dx.doi.org/10.1088/1751-8113/41/35/352002},
year = {2008},
month = {jul},
publisher = {},
volume = {41},
number = {35},
pages = {352002},
author = {Batchelor, M T and Guan, X-W and Kundu, A},
title = {{One-dimensional anyons with competing $\delta$-function and derivative $\delta$-function potentials}},
journal = {Journal of Physics A: Mathematical and Theoretical}
}

@article{HZC08,
  title = {{Ground-state properties of one-dimensional anyon gases}},
  author = {Hao, Yajiang and Zhang, Yunbo and Chen, Shu},
  journal = {Phys. Rev. A},
  volume = {78},
  issue = {2},
  pages = {023631},
  numpages = {6},
  year = {2008},
  month = {Aug},
  publisher = {American Physical Society},
  doi = {10.1103/PhysRevA.78.023631},
  url = {https://link.aps.org/doi/10.1103/PhysRevA.78.023631}
}

@article{CS09,
doi = {10.1088/1742-5468/2009/03/P03002},
url = {https://dx.doi.org/10.1088/1742-5468/2009/03/P03002},
year = {2009},
month = {mar},
publisher = {},
volume = {2009},
number = {03},
pages = {P03002},
author = {Calabrese, Pasquale and Santachiara, Raoul},
title = {{Off-diagonal correlations in one-dimensional anyonic models: a replica approach}},
journal = {Journal of Statistical Mechanics: Theory and Experiment}
}

@article{Grei09,
  title = {{Statistical phases and momentum spacings for one-dimensional anyons}},
  author = {Greiter, Martin},
  journal = {Phys. Rev. B},
  volume = {79},
  issue = {6},
  pages = {064409},
  numpages = {5},
  year = {2009},
  month = {Feb},
  publisher = {American Physical Society},
  doi = {10.1103/PhysRevB.79.064409},
  url = {https://link.aps.org/doi/10.1103/PhysRevB.79.064409}
}

@article{BS12,
  title = {{Condensation of Anyons in Frustrated Quantum Magnets}},
  author = {Batista, C. D. and Somma, Rolando D.},
  journal = {Phys. Rev. Lett.},
  volume = {109},
  issue = {22},
  pages = {227203},
  numpages = {5},
  year = {2012},
  month = {Nov},
  publisher = {American Physical Society},
  doi = {10.1103/PhysRevLett.109.227203},
  url = {https://link.aps.org/doi/10.1103/PhysRevLett.109.227203}
}

@article{WRDK14,
  title = {{Nonequilibrium Dynamics of One-Dimensional Hard-Core Anyons Following a Quench: Complete Relaxation of One-Body Observables}},
  author = {Wright, Tod M. and Rigol, Marcos and Davis, Matthew J. and Kheruntsyan, Kar\'en V.},
  journal = {Phys. Rev. Lett.},
  volume = {113},
  issue = {5},
  pages = {050601},
  numpages = {5},
  year = {2014},
  month = {Jul},
  publisher = {American Physical Society},
  doi = {10.1103/PhysRevLett.113.050601},
  url = {https://link.aps.org/doi/10.1103/PhysRevLett.113.050601}
}

@article{Zinn15,
  title = {{Strongly interacting mesoscopic systems of anyons in one dimension}},
  author = {Zinner, N. T.},
  journal = {Phys. Rev. A},
  volume = {92},
  issue = {6},
  pages = {063634},
  numpages = {6},
  year = {2015},
  month = {Dec},
  publisher = {American Physical Society},
  doi = {10.1103/PhysRevA.92.063634},
  url = {https://link.aps.org/doi/10.1103/PhysRevA.92.063634}
}

@article{MPC16,
doi = {10.1088/1742-5468/2016/07/073106},
url = {https://dx.doi.org/10.1088/1742-5468/2016/07/073106},
year = {2016},
month = {jul},
publisher = {IOP Publishing and SISSA},
volume = {2016},
number = {7},
pages = {073106},
author = {Marmorini, Giacomo and Pepe, Michele and Calabrese, Pasquale},
title = {One-body reduced density matrix of trapped impenetrable anyons in one dimension},
journal = {Journal of Statistical Mechanics: Theory and Experiment}
}

@article{Hao16,
  title = {{Ground-state properties of hard-core anyons in a harmonic potential}},
  author = {Hao, Yajiang},
  journal = {Phys. Rev. A},
  volume = {93},
  issue = {6},
  pages = {063627},
  numpages = {6},
  year = {2016},
  month = {Jun},
  publisher = {American Physical Society},
  doi = {10.1103/PhysRevA.93.063627},
  url = {https://link.aps.org/doi/10.1103/PhysRevA.93.063627}
}

@article{PC17,
  title = {{Exact dynamics following an interaction quench in a one-dimensional anyonic gas}},
  author = {Piroli, Lorenzo and Calabrese, Pasquale},
  journal = {Phys. Rev. A},
  volume = {96},
  issue = {2},
  pages = {023611},
  numpages = {22},
  year = {2017},
  month = {Aug},
  publisher = {American Physical Society},
  doi = {10.1103/PhysRevA.96.023611},
  url = {https://link.aps.org/doi/10.1103/PhysRevA.96.023611}
}

@article{RCCM19,
  title = {{Anyonic tight-binding models of parafermions and of fractionalized fermions}},
  author = {Rossini, Davide and Carrega, Matteo and Calvanese Strinati, Marcello and Mazza, Leonardo},
  journal = {Phys. Rev. B},
  volume = {99},
  issue = {8},
  pages = {085113},
  numpages = {12},
  year = {2019},
  month = {Feb},
  publisher = {American Physical Society},
  doi = {10.1103/PhysRevB.99.085113},
  url = {https://link.aps.org/doi/10.1103/PhysRevB.99.085113}
}

@article{SPC20,
doi = {10.1088/1742-5468/abaed1},
url = {https://dx.doi.org/10.1088/1742-5468/abaed1},
year = {2020},
month = {sep},
publisher = {IOP Publishing and SISSA},
volume = {2020},
number = {9},
pages = {093103},
author = {Scopa, Stefano and Piroli, Lorenzo and Calabrese, Pasquale},
title = {{One-particle density matrix of a trapped Lieb–Liniger anyonic gas}},
journal = {Journal of Statistical Mechanics: Theory and Experiment}
}

@article{HK20,
  author    = {N. L. Harshman and A. C. Knapp},
  title     = {{Anyons from three-body hard-core interactions in one dimension}},
  journal   = {Annals of Physics},
  volume    = {413},
  pages     = {168065},
  year      = {2020},
  doi       = {10.1016/j.aop.2019.168065}
}

@article{MD21,
  title = {{Thermodynamics of Statistical Anyons}},
  author = {Myers, Nathan M. and Deffner, Sebastian},
  journal = {PRX Quantum},
  volume = {2},
  issue = {4},
  pages = {040312},
  numpages = {28},
  year = {2021},
  month = {Oct},
  publisher = {American Physical Society},
  doi = {10.1103/PRXQuantum.2.040312},
  url = {https://link.aps.org/doi/10.1103/PRXQuantum.2.040312}
}

@article{HK22,
  title = {{Topological exchange statistics in one dimension}},
  author = {Harshman, N. L. and Knapp, A. C.},
  journal = {Phys. Rev. A},
  volume = {105},
  issue = {5},
  pages = {052214},
  numpages = {15},
  year = {2022},
  month = {May},
  publisher = {American Physical Society},
  doi = {10.1103/PhysRevA.105.052214},
  url = {https://link.aps.org/doi/10.1103/PhysRevA.105.052214}
}

@article{Patu23b,
  title = {{Nonequilibrium dynamics in one-dimensional strongly interacting two-component gases}},
  author = {P\^{a}\c{t}u, Ovidiu I.},
  journal = {Phys. Rev. A},
  volume = {108},
  issue = {5},
  pages = {053304},
  numpages = {27},
  year = {2023},
  month = {Nov},
  publisher = {American Physical Society},
  doi = {10.1103/PhysRevA.108.053304},
  url = {https://link.aps.org/doi/10.1103/PhysRevA.108.053304}
}

@article{MYC23,
  doi = {10.22331/q-2023-12-20-1211},
  url = {https://doi.org/10.22331/q-2023-12-20-1211},
  title = {Quantum {A}lchemy and {U}niversal {O}rthogonality {C}atastrophe in {O}ne-{D}imensional {A}nyons},
  author = {Mackel, Naim E. and Yang, Jing and Campo, Adolfo del},
  journal = {{Quantum}},
  issn = {2521-327X},
  publisher = {{Verein zur F{\"{o}}rderung des Open Access Publizierens in den Quantenwissenschaften}},
  volume = {7},
  pages = {1211},
  month = dec,
  year = {2023}
}

@article{MWS20,
	title={{Phase diagram of the $\mathbb{Z}_3$-Fock parafermion chain with pair hopping}},
	author={Iman Mahyaeh and Jurriaan Wouters and Dirk Schuricht},
	journal={SciPost Phys. Core},
	volume={3},
	pages={011},
	year={2020},
	publisher={SciPost},
	doi={10.21468/SciPostPhysCore.3.2.011},
	url={https://scipost.org/10.21468/SciPostPhysCore.3.2.011},
}

@article{BJEP21,
  title = {{Bosonic Continuum Theory of One-Dimensional Lattice Anyons}},
  author = {Bonkhoff, Martin and J\"agering, Kevin and Eggert, Sebastian and Pelster, Axel and Thorwart, Michael and Posske, Thore},
  journal = {Phys. Rev. Lett.},
  volume = {126},
  issue = {16},
  pages = {163201},
  numpages = {7},
  year = {2021},
  month = {Apr},
  publisher = {American Physical Society},
  doi = {10.1103/PhysRevLett.126.163201},
  url = {https://link.aps.org/doi/10.1103/PhysRevLett.126.163201}
}

@article{AN07,
  title = {{Coulomb Blockade of Anyons in Quantum Antidots}},
  author = {Averin, Dmitri V. and Nesteroff, James A.},
  journal = {Phys. Rev. Lett.},
  volume = {99},
  issue = {9},
  pages = {096801},
  numpages = {4},
  year = {2007},
  month = {Aug},
  publisher = {American Physical Society},
  doi = {10.1103/PhysRevLett.99.096801},
  url = {https://link.aps.org/doi/10.1103/PhysRevLett.99.096801}
}

@article{HZC09,
  title = {{Ground-state properties of hard-core anyons in one-dimensional optical lattices}},
  author = {Hao, Yajiang and Zhang, Yunbo and Chen, Shu},
  journal = {Phys. Rev. A},
  volume = {79},
  issue = {4},
  pages = {043633},
  numpages = {5},
  year = {2009},
  month = {Apr},
  publisher = {American Physical Society},
  doi = {10.1103/PhysRevA.79.043633},
  url = {https://link.aps.org/doi/10.1103/PhysRevA.79.043633}
}

@article{BCM09,
  title = {{Junctions of anyonic Luttinger wires}},
  author = {Bellazzini, Brando and Calabrese, Pasquale and Mintchev, Mihail},
  journal = {Phys. Rev. B},
  volume = {79},
  issue = {8},
  pages = {085122},
  numpages = {15},
  year = {2009},
  month = {Feb},
  publisher = {American Physical Society},
  doi = {10.1103/PhysRevB.79.085122},
  url = {https://link.aps.org/doi/10.1103/PhysRevB.79.085122}
}

@Article{KLMR11,
author={Keilmann, Tassilo
and Lanzmich, Simon
and McCulloch, Ian
and Roncaglia, Marco},
title={{Statistically induced phase transitions and anyons in 1D optical lattices}},
journal={Nature Communications},
year={2011},
month={Jun},
day={21},
volume={2},
number={1},
pages={361},
issn={2041-1723},
doi={10.1038/ncomms1353},
url={https://doi.org/10.1038/ncomms1353}
}

@article{LD12,
  title = {{Anyonic Bloch oscillations}},
  author = {Longhi, Stefano and Della Valle, Giuseppe},
  journal = {Phys. Rev. B},
  volume = {85},
  issue = {16},
  pages = {165144},
  numpages = {8},
  year = {2012},
  month = {Apr},
  publisher = {American Physical Society},
  doi = {10.1103/PhysRevB.85.165144},
  url = {https://link.aps.org/doi/10.1103/PhysRevB.85.165144}
}

@article{WWZ14,
  title = {{Quantum walks of two interacting anyons in one-dimensional optical lattices}},
  author = {Wang, Limin and Wang, Li and Zhang, Yunbo},
  journal = {Phys. Rev. A},
  volume = {90},
  issue = {6},
  pages = {063618},
  numpages = {6},
  year = {2014},
  month = {Dec},
  publisher = {American Physical Society},
  doi = {10.1103/PhysRevA.90.063618},
  url = {https://link.aps.org/doi/10.1103/PhysRevA.90.063618}
}

@article{TEP15,
doi = {10.1088/1367-2630/17/12/123016},
url = {https://dx.doi.org/10.1088/1367-2630/17/12/123016},
year = {2015},
month = {dec},
publisher = {IOP Publishing},
volume = {17},
number = {12},
pages = {123016},
author = {Tang, Guixin and Eggert, Sebastian and Pelster, Axel},
title = {{Ground-state properties of anyons in a one-dimensional lattice}},
journal = {New Journal of Physics}
}

@article{RFB14,
  title = {{Anyonic Liquids in Nearly Saturated Spin Chains}},
  author = {Rahmani, Armin and Feiguin, Adrian E. and Batista, Cristian D.},
  journal = {Phys. Rev. Lett.},
  volume = {113},
  issue = {26},
  pages = {267201},
  numpages = {6},
  year = {2014},
  month = {Dec},
  publisher = {American Physical Society},
  doi = {10.1103/PhysRevLett.113.267201},
  url = {https://link.aps.org/doi/10.1103/PhysRevLett.113.267201}
}

@article{GS15,
  title = {{Anyon Hubbard Model in One-Dimensional Optical Lattices}},
  author = {Greschner, Sebastian and Santos, Luis},
  journal = {Phys. Rev. Lett.},
  volume = {115},
  issue = {5},
  pages = {053002},
  numpages = {5},
  year = {2015},
  month = {Jul},
  publisher = {American Physical Society},
  doi = {10.1103/PhysRevLett.115.053002},
  url = {https://link.aps.org/doi/10.1103/PhysRevLett.115.053002}
}

@article{AFS16,
  title = {{Critical points of the anyon-Hubbard model}},
  author = {Arcila-Forero, J. and Franco, R. and Silva-Valencia, J.},
  journal = {Phys. Rev. A},
  volume = {94},
  issue = {1},
  pages = {013611},
  numpages = {9},
  year = {2016},
  month = {Jul},
  publisher = {American Physical Society},
  doi = {10.1103/PhysRevA.94.013611},
  url = {https://link.aps.org/doi/10.1103/PhysRevA.94.013611}
}

@article{ZGFS17,
  title = {{Ground-state properties of the one-dimensional unconstrained pseudo-anyon Hubbard model}},
  author = {Zhang, Wanzhou and Greschner, Sebastian and Fan, Ernv and Scott, Tony C. and Zhang, Yunbo},
  journal = {Phys. Rev. A},
  volume = {95},
  issue = {5},
  pages = {053614},
  numpages = {10},
  year = {2017},
  month = {May},
  publisher = {American Physical Society},
  doi = {10.1103/PhysRevA.95.053614},
  url = {https://link.aps.org/doi/10.1103/PhysRevA.95.053614}
}

@article{LGDG18,
  title = {{Asymmetric Particle Transport and Light-Cone Dynamics Induced by Anyonic Statistics}},
  author = {Liu, Fangli and Garrison, James R. and Deng, Dong-Ling and Gong, Zhe-Xuan and Gorshkov, Alexey V.},
  journal = {Phys. Rev. Lett.},
  volume = {121},
  issue = {25},
  pages = {250404},
  numpages = {7},
  year = {2018},
  month = {Dec},
  publisher = {American Physical Society},
  doi = {10.1103/PhysRevLett.121.250404},
  url = {https://link.aps.org/doi/10.1103/PhysRevLett.121.250404}
}

@Article{GIZ21,
	title={{Effective free-fermionic form factors and the XY spin chain}},
	author={O. Gamayun and N. Iorgov and Yu. Zhuravlev},
	journal={SciPost Phys.},
	volume={10},
	pages={070},
	year={2021},
	publisher={SciPost},
	doi={10.21468/SciPostPhys.10.3.070},
	url={https://scipost.org/10.21468/SciPostPhys.10.3.070},
}

@article{TGB21,
  title = {{Gaussian optical networks for one-dimensional anyons}},
  author = {Tosta, Allan D. C. and Galv\~{a}o, Ernesto F. and Brod, Daniel J.},
  journal = {Phys. Rev. A},
  volume = {104},
  issue = {2},
  pages = {022604},
  numpages = {13},
  year = {2021},
  month = {Aug},
  publisher = {American Physical Society},
  doi = {10.1103/PhysRevA.104.022604},
  url = {https://link.aps.org/doi/10.1103/PhysRevA.104.022604}
}

@article{ZNIG22,
  title = {{Large-time and long-distance asymptotics of the thermal correlators of the impenetrable anyonic lattice gas}},
  author = {Zhuravlev, Yuri and Naichuk, Eduard and Iorgov, Nikolai and Gamayun, Oleksandr},
  journal = {Phys. Rev. B},
  volume = {105},
  issue = {8},
  pages = {085145},
  numpages = {12},
  year = {2022},
  month = {Feb},
  publisher = {American Physical Society},
  doi = {10.1103/PhysRevB.105.085145},
  url = {https://link.aps.org/doi/10.1103/PhysRevB.105.085145}
}

@Article{CG22,
	title={{On the dynamics of free-fermionic tau-functions at finite temperature}},
	author={Daniel Chernowitz and Oleksandr Gamayun},
	journal={SciPost Phys. Core},
	volume={5},
	pages={006},
	year={2022},
	publisher={SciPost},
	doi={10.21468/SciPostPhysCore.5.1.006},
	url={https://scipost.org/10.21468/SciPostPhysCore.5.1.006},
}

@article{Wang22,
  title = {{Exact dynamical correlations of hard-core anyons in one-dimensional lattices}},
  author = {Wang, Qing-Wei},
  journal = {Phys. Rev. B},
  volume = {105},
  issue = {20},
  pages = {205143},
  numpages = {11},
  year = {2022},
  month = {May},
  publisher = {American Physical Society},
  doi = {10.1103/PhysRevB.105.205143},
  url = {https://link.aps.org/doi/10.1103/PhysRevB.105.205143}
}

@article{TLBI24,
  title = {{Fermionic anyons: Entanglement and quantum computation from a resource-theoretic perspective}},
  author = {Tosta, Allan and Louren\c{c}o, Ant\^{o}nio C. and Brod, Daniel and Iemini, Fernando and Debarba, Tiago},
  journal = {Phys. Rev. A},
  volume = {110},
  issue = {1},
  pages = {L010404},
  numpages = {6},
  year = {2024},
  month = {Jul},
  publisher = {American Physical Society},
  doi = {10.1103/PhysRevA.110.L010404},
  url = {https://link.aps.org/doi/10.1103/PhysRevA.110.L010404}
}

@article{NWKE24,
	title={{Beyond braid statistics: Constructing a lattice model for anyons with exchange statistics intrinsic to one dimension}},
	author={Sebastian Nagies and Botao Wang and Adam C. Knapp and André Eckardt and Nathan L. Harshman},
	journal={SciPost Phys.},
	volume={16},
	pages={086},
	year={2024},
	publisher={SciPost},
	doi={10.21468/SciPostPhys.16.3.086},
	url={https://scipost.org/10.21468/SciPostPhys.16.3.086},
}

@article{PKF24,
  title = {{Exact spectral function and nonequilibrium dynamics of the strongly interacting Hubbard model}},
  author = {P\^{a}\c{t}u, Ovidiu I. and Kl\"umper, Andreas and Foerster, Angela},
  journal = {Phys. Rev. B},
  volume = {110},
  issue = {20},
  pages = {205101},
  numpages = {9},
  year = {2024},
  month = {Nov},
  publisher = {American Physical Society},
  doi = {10.1103/PhysRevB.110.205101},
  url = {https://link.aps.org/doi/10.1103/PhysRevB.110.205101}
}

@article{BJHP25,
  title = {{Anyonic Phase Transitions in the 1D Extended Hubbard Model with Fractional Statistics}},
  author = {Bonkhoff, Martin and J\"{a}gering, Kevin and Hu, Shijie and Pelster, Axel and Eggert, Sebastian and Schneider, Imke},
  journal = {Phys. Rev. Lett.},
  volume = {135},
  issue = {3},
  pages = {036601},
  numpages = {10},
  year = {2025},
  month = {Jul},
  publisher = {American Physical Society},
  doi = {10.1103/7n1c-vq2p},
  url = {https://link.aps.org/doi/10.1103/7n1c-vq2p}
}

@Article{ZYWD22,
author={Zhang, Weixuan
and Yuan, Hao
and Wang, Haiteng
and Di, Fengxiao
and Sun, Na
and Zheng, Xingen
and Sun, Houjun
and Zhang, Xiangdong},
title={{Observation of Bloch oscillations dominated by effective anyonic particle statistics}},
journal={Nature Communications},
year={2022},
month={May},
day={02},
volume={13},
number={1},
pages={2392},
issn={2041-1723},
doi={10.1038/s41467-022-29895-0},
url={https://doi.org/10.1038/s41467-022-29895-0}
}

@article{KSLK24,
author = {Joyce Kwan  and Perrin Segura  and Yanfei Li  and Sooshin Kim  and Alexey V. Gorshkov  and André Eckardt  and Brice Bakkali-Hassani  and Markus Greiner },
title = {{Realization of one-dimensional anyons with arbitrary statistical phase}},
journal = {Science},
volume = {386},
number = {6725},
pages = {1055-1060},
year = {2024},
doi = {10.1126/science.adi3252},
URL = {https://www.science.org/doi/abs/10.1126/science.adi3252}
}

@Article{DWHV25,
author={Dhar, Sudipta
and Wang, Botao
and Horvath, Milena
and Vashisht, Amit
and Zeng, Yi
and Zvonarev, Mikhail B.
and Goldman, Nathan
and Guo, Yanliang
and Landini, Manuele
and N{\"a}gerl, Hanns-Christoph},
title={{Observing anyonization of bosons in a quantum gas}},
journal={Nature},
year={2025},
month={Jun},
day={01},
volume={642},
number={8066},
pages={53-57},
issn={1476-4687},
doi={10.1038/s41586-025-09016-9},
url={https://doi.org/10.1038/s41586-025-09016-9}
}

@article{Gira60,
    author = {Girardeau, M.},
    title = {{Relationship between Systems of Impenetrable Bosons and Fermions in One Dimension}},
    journal = {Journal of Mathematical Physics},
    volume = {1},
    number = {6},
    pages = {516-523},
    year = {1960},
    month = {11},
    issn = {0022-2488},
    doi = {10.1063/1.1703687},
    url = {https://doi.org/10.1063/1.1703687}
}

@article{CS98,
title = {Realizing discontinuous wave functions with renormalized short-range potentials},
journal = {Physics Letters A},
volume = {243},
number = {3},
pages = {111-116},
year = {1998},
issn = {0375-9601},
doi = {https://doi.org/10.1016/S0375-9601(98)00188-1},
url = {https://www.sciencedirect.com/science/article/pii/S0375960198001881},
author = {Taksu Cheon and T Shigehara}
}

@article{CS99,
  title = {{Fermion-Boson Duality of One-Dimensional Quantum Particles with Generalized Contact Interactions}},
  author = {Cheon, Taksu and Shigehara, T.},
  journal = {Phys. Rev. Lett.},
  volume = {82},
  issue = {12},
  pages = {2536--2539},
  numpages = {0},
  year = {1999},
  month = {Mar},
  publisher = {American Physical Society},
  doi = {10.1103/PhysRevLett.82.2536},
  url = {https://link.aps.org/doi/10.1103/PhysRevLett.82.2536}
}

@article{PP70,
author = { Popov, V S  and  Perelomov, A M},
    title = {{Parametric excitation of a quantum oscillator II}},
    journal = {Zh. Eksp. Teor. Fiz.},
    volume = {57},
    number = {},
    pages = {1684},
    year = {1970},
    url = {http://jetp.ras.ru/cgi-bin/dn/e_030_05_0910.pdf}
}

@book{PZ98,
  title     = {Quantum Mechanics: Selected Topics},
  author    = {A. M. Perelomov and Y. B. Zel'dovich},
  year      = {1998},
  publisher = {World Scientific},
  address   = {Singapore}
}

@article{GB04,
  title = {{Tuning the Interactions of Spin-Polarized Fermions Using Quasi-One-Dimensional Confinement}},
  author = {Granger, Brian E. and Blume, D.},
  journal = {Phys. Rev. Lett.},
  volume = {92},
  issue = {13},
  pages = {133202},
  numpages = {4},
  year = {2004},
  month = {Apr},
  publisher = {American Physical Society},
  doi = {10.1103/PhysRevLett.92.133202},
  url = {https://link.aps.org/doi/10.1103/PhysRevLett.92.133202}
}

@article{GO04,
  title = {{Theory of spinor Fermi and Bose gases in tight atom waveguides}},
  author = {Girardeau, M. D. and Olshanii, M.},
  journal = {Phys. Rev. A},
  volume = {70},
  issue = {2},
  pages = {023608},
  numpages = {4},
  year = {2004},
  month = {Aug},
  publisher = {American Physical Society},
  doi = {10.1103/PhysRevA.70.023608},
  url = {https://link.aps.org/doi/10.1103/PhysRevA.70.023608}
}

@article{BEG05,
  title = {{Exponentially Decaying Correlations in a Gas of Strongly Interacting Spin-Polarized 1D Fermions with Zero-Range Interactions}},
  author = {Bender, Scott A. and Erker, Kevin D. and Granger, Brian E.},
  journal = {Phys. Rev. Lett.},
  volume = {95},
  issue = {23},
  pages = {230404},
  numpages = {4},
  year = {2005},
  month = {Nov},
  publisher = {American Physical Society},
  doi = {10.1103/PhysRevLett.95.230404},
  url = {https://link.aps.org/doi/10.1103/PhysRevLett.95.230404}
}

@article{GM06,
  title = {{Bosonization, Pairing, and Superconductivity of the Fermionic Tonks-Girardeau Gas}},
  author = {Girardeau, M. D. and Minguzzi, A.},
  journal = {Phys. Rev. Lett.},
  volume = {96},
  issue = {8},
  pages = {080404},
  numpages = {4},
  year = {2006},
  month = {Mar},
  publisher = {American Physical Society},
  doi = {10.1103/PhysRevLett.96.080404},
  url = {https://link.aps.org/doi/10.1103/PhysRevLett.96.080404}
}

@article{HZC07,
  title = {{One-dimensional fermionic gases with attractive $p$-wave interaction in a hard-wall trap}},
  author = {Hao, Yajiang and Zhang, Yunbo and Chen, Shu},
  journal = {Phys. Rev. A},
  volume = {76},
  issue = {6},
  pages = {063601},
  numpages = {6},
  year = {2007},
  month = {Dec},
  publisher = {American Physical Society},
  doi = {10.1103/PhysRevA.76.063601},
  url = {https://link.aps.org/doi/10.1103/PhysRevA.76.063601}
}

@article{Pric08,
  title = {{Resonant Scattering of Ultracold Atoms in Low Dimensions}},
  author = {Pricoupenko, Ludovic},
  journal = {Phys. Rev. Lett.},
  volume = {100},
  issue = {17},
  pages = {170404},
  numpages = {4},
  year = {2008},
  month = {May},
  publisher = {American Physical Society},
  doi = {10.1103/PhysRevLett.100.170404},
  url = {https://link.aps.org/doi/10.1103/PhysRevLett.100.170404}
}

@article{ILGG10,
  title = {{Exact Solution for 1D Spin-Polarized Fermions with Resonant Interactions}},
  author = {Imambekov, Adilet and Lukyanov, Alexander A. and Glazman, Leonid I. and Gritsev, Vladimir},
  journal = {Phys. Rev. Lett.},
  volume = {104},
  issue = {4},
  pages = {040402},
  numpages = {4},
  year = {2010},
  month = {Jan},
  publisher = {American Physical Society},
  doi = {10.1103/PhysRevLett.104.040402},
  url = {https://link.aps.org/doi/10.1103/PhysRevLett.104.040402}
}

@article{Cui16,
  title = {{Universal one-dimensional atomic gases near odd-wave resonance}},
  author = {Cui, Xiaoling},
  journal = {Phys. Rev. A},
  volume = {94},
  issue = {4},
  pages = {043636},
  numpages = {9},
  year = {2016},
  month = {Oct},
  publisher = {American Physical Society},
  doi = {10.1103/PhysRevA.94.043636},
  url = {https://link.aps.org/doi/10.1103/PhysRevA.94.043636}
}

@article{YGC16,
  title = {{Engineering quantum magnetism in one-dimensional trapped Fermi gases with $p$-wave interactions}},
  author = {Yang, Lijun and Guan, Xiwen and Cui, Xiaoling},
  journal = {Phys. Rev. A},
  volume = {93},
  issue = {5},
  pages = {051605},
  numpages = {5},
  year = {2016},
  month = {May},
  publisher = {American Physical Society},
  doi = {10.1103/PhysRevA.93.051605},
  url = {https://link.aps.org/doi/10.1103/PhysRevA.93.051605}
}

@article{CLH16,
  title = {{Probing an effective-range-induced super fermionic Tonks-Girardeau gas with ultracold atoms in one-dimensional harmonic traps}},
  author = {Chen, Xiao-Long and Liu, Xia-Ji and Hu, Hui},
  journal = {Phys. Rev. A},
  volume = {94},
  issue = {3},
  pages = {033630},
  numpages = {6},
  year = {2016},
  month = {Sep},
  publisher = {American Physical Society},
  doi = {10.1103/PhysRevA.94.033630},
  url = {https://link.aps.org/doi/10.1103/PhysRevA.94.033630}
}

@article{PCC18,
  title = {{Many-body stabilization of a resonant $p$-wave Fermi gas in one dimension}},
  author = {Pan, Lei and Chen, Shu and Cui, Xiaoling},
  journal = {Phys. Rev. A},
  volume = {98},
  issue = {1},
  pages = {011603},
  numpages = {5},
  year = {2018},
  month = {Jul},
  publisher = {American Physical Society},
  doi = {10.1103/PhysRevA.98.011603},
  url = {https://link.aps.org/doi/10.1103/PhysRevA.98.011603}
}

@article{STN18,
  title = {{Comparative study of one-dimensional Bose and Fermi gases with contact interactions from the viewpoint of universal relations for correlation functions}},
  author = {Sekino, Yuta and Tan, Shina and Nishida, Yusuke},
  journal = {Phys. Rev. A},
  volume = {97},
  issue = {1},
  pages = {013621},
  numpages = {7},
  year = {2018},
  month = {Jan},
  publisher = {American Physical Society},
  doi = {10.1103/PhysRevA.97.013621},
  url = {https://link.aps.org/doi/10.1103/PhysRevA.97.013621}
}

@article{XGZS18,
  title = {{Momentum distribution and contacts of one-dimensional spinless Fermi gases with an attractive $p$-wave interaction}},
  author = {Yin, Xiangguo and Guan, Xi-Wen and Zhang, Yunbo and Su, Haibin and Zhang, Shizhong},
  journal = {Phys. Rev. A},
  volume = {98},
  issue = {2},
  pages = {023605},
  numpages = {14},
  year = {2018},
  month = {Aug},
  publisher = {American Physical Society},
  doi = {10.1103/PhysRevA.98.023605},
  url = {https://link.aps.org/doi/10.1103/PhysRevA.98.023605}
}

@article{KS20,
doi = {10.1088/1367-2630/abb386},
url = {https://dx.doi.org/10.1088/1367-2630/abb386},
year = {2020},
month = {sep},
publisher = {IOP Publishing},
volume = {22},
number = {9},
pages = {093053},
author = {Ko\'{s}cik, Przemys\l{l}aw and Sowi\'{n}ski, Tomasz},
title = {{Variational ansatz for p-wave fermions confined in a one-dimensional harmonic trap}},
journal = {New Journal of Physics}
}

@article{KS23,
  title = {{Universality of Internal Correlations of Strongly Interacting $p$-Wave Fermions in One-Dimensional Geometry}},
  author = {Ko\'{s}cik, Przemys\l{l}aw and Sowi\'{n}ski, Tomasz},
  journal = {Phys. Rev. Lett.},
  volume = {130},
  issue = {25},
  pages = {253401},
  numpages = {5},
  year = {2023},
  month = {Jun},
  publisher = {American Physical Society},
  doi = {10.1103/PhysRevLett.130.253401},
  url = {https://link.aps.org/doi/10.1103/PhysRevLett.130.253401}
}

@article{SRAJ24,
  title = {{Universal Composite Boson Formation in Strongly Interacting One-Dimensional Fermionic Systems}},
  author = {Sabater, Francesc and Rojo-Franc\`{a}s, Abel and Astrakharchik, Grigori E. and Juli\'{a}-D\'{i}az, Bruno},
  journal = {Phys. Rev. Lett.},
  volume = {132},
  issue = {19},
  pages = {193401},
  numpages = {5},
  year = {2024},
  month = {May},
  publisher = {American Physical Society},
  doi = {10.1103/PhysRevLett.132.193401},
  url = {https://link.aps.org/doi/10.1103/PhysRevLett.132.193401}
}

@article{SRAJ25,
  title = {{Bardeen-Cooper-Schrieffer state representation and pairing detection in the fermionic Tonks-Girardeau gas}},
  author = {Sabater, Francesc and Rojo-Franc\`{a}s, Abel and Astrakharchik, Grigori E. and Juli\'{a}-D\'{i}az, Bruno},
  journal = {Phys. Rev. Res.},
  volume = {7},
  issue = {3},
  pages = {L032007},
  numpages = {6},
  year = {2025},
  month = {Jul},
  publisher = {American Physical Society},
  doi = {10.1103/cqkw-bf3v},
  url = {https://link.aps.org/doi/10.1103/cqkw-bf3v}
}

@article{WCC25,
  title = {{Boson-anyon-fermion mapping and anyon construction in one dimension}},
  author = {Wang, Haitian and Chen, Yu and Cui, Xiaoling},
  journal = {Phys. Rev. Res.},
  volume = {7},
  issue = {2},
  pages = {L022075},
  numpages = {5},
  year = {2025},
  month = {Jun},
  publisher = {American Physical Society},
  doi = {10.1103/np63-xnh8},
  url = {https://link.aps.org/doi/10.1103/np63-xnh8}
}

@article{HBB25a,
  author = {Hidalgo-Sacoto, R and Busch, T. and  Blume, D.},
  title  = {{Universal momentum tail of identical one-dimensional anyons with two-body interactions}},
  journal = {arXiv:2505.17669},
  year = {2025},
  url = {https://arxiv.org/pdf/2505.17669}
}

@article{HBB25b,
  author = {Hidalgo-Sacoto, R and Busch, T. and  Blume, D.},
  title  = {{Two identical 1D anyons with zero-range interactions: Exchange statistics, scattering theory, and anyon-anyon mapping}},
  journal = {arXiv:2505.23127},
  year = {2025},
  url = {https://arxiv.org/pdf/2505.23127}
}

@article{LL63,
  title = {{Exact Analysis of an Interacting {Bose} Gas. I. The General Solution and the Ground State}},
  author = {Lieb, Elliott H. and Liniger, Werner},
  journal = {Phys. Rev.},
  volume = {130},
  issue = {},
  pages = {1605--1616},
  numpages = {0},
  year = {1963},
  month = {},
  publisher = {American Physical Society},
  doi = {10.1103/PhysRev.130.1605},
  url = {https://link.aps.org/doi/10.1103/PhysRev.130.1605}
}

@article{YTZ15,
  title = {{Universal Relations for a Fermi Gas Close to a $p$-Wave Interaction Resonance}},
  author = {Yu, Zhenhua and Thywissen, Joseph H. and Zhang, Shizhong},
  journal = {Phys. Rev. Lett.},
  volume = {115},
  issue = {13},
  pages = {135304},
  numpages = {5},
  year = {2015},
  month = {Sep},
  publisher = {American Physical Society},
  doi = {10.1103/PhysRevLett.115.135304},
  url = {https://link.aps.org/doi/10.1103/PhysRevLett.115.135304}
}

@article{YU15,
  title = {Universal High-Momentum Asymptote and Thermodynamic Relations in a Spinless Fermi Gas with a Resonant $p$-Wave Interaction},
  author = {Yoshida, Shuhei M. and Ueda, Masahito},
  journal = {Phys. Rev. Lett.},
  volume = {115},
  issue = {13},
  pages = {135303},
  numpages = {5},
  year = {2015},
  month = {Sep},
  publisher = {American Physical Society},
  doi = {10.1103/PhysRevLett.115.135303},
  url = {https://link.aps.org/doi/10.1103/PhysRevLett.115.135303}
}

@Article{LTSY16,
author={Luciuk, Christopher
and Trotzky, Stefan
and Smale, Scott
and Yu, Zhenhua
and Zhang, Shizhong
and Thywissen, Joseph H.},
title={{Evidence for universal relations describing a gas with p-wave interactions}},
journal={Nature Physics},
year={2016},
month={Jun},
volume={12},
number={6},
pages={599-605},
doi={10.1038/nphys3670},
url={https://doi.org/10.1038/nphys3670}
}

@article{HZCZ16,
  title = {Concept of a Contact Spectrum and Its Applications in Atomic Quantum Hall States},
  author = {He, Mingyuan and Zhang, Shaoliang and Chan, Hon Ming and Zhou, Qi},
  journal = {Phys. Rev. Lett.},
  volume = {116},
  issue = {4},
  pages = {045301},
  numpages = {5},
  year = {2016},
  month = {Jan},
  publisher = {American Physical Society},
  doi = {10.1103/PhysRevLett.116.045301},
  url = {https://link.aps.org/doi/10.1103/PhysRevLett.116.045301}
}

@article{PLH16,
  title = {{Large-momentum distribution of a polarized Fermi gas and $p$-wave contacts}},
  author = {Peng, Shi-Guo and Liu, Xia-Ji and Hu, Hui},
  journal = {Phys. Rev. A},
  volume = {94},
  issue = {6},
  pages = {063651},
  numpages = {8},
  year = {2016},
  month = {Dec},
  publisher = {American Physical Society},
  doi = {10.1103/PhysRevA.94.063651},
  url = {https://link.aps.org/doi/10.1103/PhysRevA.94.063651}
}

@article{QCY16,
  title = {Universal relations and normal phase of an ultracold Fermi gas with coexisting $s$- and $p$-wave interactions},
  author = {Qin, Fang and Cui, Xiaoling and Yi, Wei},
  journal = {Phys. Rev. A},
  volume = {94},
  issue = {6},
  pages = {063616},
  numpages = {9},
  year = {2016},
  month = {Dec},
  publisher = {American Physical Society},
  doi = {10.1103/PhysRevA.94.063616},
  url = {https://link.aps.org/doi/10.1103/PhysRevA.94.063616}
}

@article{ZHZ17,
  title = {{Contact matrix in dilute quantum systems}},
  author = {Zhang, Shao-Liang and He, Mingyuan and Zhou, Qi},
  journal = {Phys. Rev. A},
  volume = {95},
  issue = {6},
  pages = {062702},
  numpages = {6},
  year = {2017},
  month = {Jun},
  publisher = {American Physical Society},
  doi = {10.1103/PhysRevA.95.062702},
  url = {https://link.aps.org/doi/10.1103/PhysRevA.95.062702}
}

@article{Qin18,
  title = {{Universal relations and normal-state properties of a Fermi gas with laser-dressed mixed-partial-wave interactions}},
  author = {Qin, Fang},
  journal = {Phys. Rev. A},
  volume = {98},
  issue = {5},
  pages = {053621},
  numpages = {13},
  year = {2018},
  month = {Nov},
  publisher = {American Physical Society},
  doi = {10.1103/PhysRevA.98.053621},
  url = {https://link.aps.org/doi/10.1103/PhysRevA.98.053621}
}

@article{HZ21,
  title = {{$p$-wave contacts of quantum gases in quasi-one-dimensional and quasi-two-dimensional traps}},
  author = {He, Mingyuan and Zhou, Qi},
  journal = {Phys. Rev. A},
  volume = {104},
  issue = {4},
  pages = {043303},
  numpages = {8},
  year = {2021},
  month = {Oct},
  publisher = {American Physical Society},
  doi = {10.1103/PhysRevA.104.043303},
  url = {https://link.aps.org/doi/10.1103/PhysRevA.104.043303}
}

@article{MLZ24,
  title = {{Universal relations for dilute systems with two-body decays in reduced dimensions}},
  author = {He, Mingyuan and Lv, Chenwei and Zhou, Qi},
  journal = {Phys. Rev. A},
  volume = {109},
  issue = {6},
  pages = {063301},
  numpages = {10},
  year = {2024},
  month = {Jun},
  publisher = {American Physical Society},
  doi = {10.1103/PhysRevA.109.063301},
  url = {https://link.aps.org/doi/10.1103/PhysRevA.109.063301}
}

@article{AG03,
  title = {{Correlation functions and momentum distribution of one-dimensional Bose systems}},
  author = {Astrakharchik, G. E. and Giorgini, S.},
  journal = {Phys. Rev. A},
  volume = {68},
  issue = {3},
  pages = {031602},
  numpages = {4},
  year = {2003},
  month = {Sep},
  publisher = {American Physical Society},
  doi = {10.1103/PhysRevA.68.031602},
  url = {https://link.aps.org/doi/10.1103/PhysRevA.68.031602}
}

@article{RM05,
  title = {{Fermionization in an Expanding 1D Gas of Hard-Core Bosons}},
  author = {Rigol, Marcos and Muramatsu, Alejandro},
  journal = {Phys. Rev. Lett.},
  volume = {94},
  issue = {24},
  pages = {240403},
  numpages = {4},
  year = {2005},
  month = {Jun},
  publisher = {American Physical Society},
  doi = {10.1103/PhysRevLett.94.240403},
  url = {https://link.aps.org/doi/10.1103/PhysRevLett.94.240403}
}

@article{MG05,
  title = {{Exact Coherent States of a Harmonically Confined Tonks-Girardeau Gas}},
  author = {Minguzzi, A. and Gangardt, D. M.},
  journal = {Phys. Rev. Lett.},
  volume = {94},
  issue = {24},
  pages = {240404},
  numpages = {4},
  year = {2005},
  month = {Jun},
  publisher = {American Physical Society},
  doi = {10.1103/PhysRevLett.94.240404},
  url = {https://link.aps.org/doi/10.1103/PhysRevLett.94.240404}
}

@article{WMLZ20,
author = {Joshua M. Wilson  and Neel Malvania  and Yuan Le  and Yicheng Zhang  and Marcos Rigol  and David S. Weiss },
title = {{Observation of dynamical fermionization}},
journal = {Science},
volume = {367},
number = {6485},
pages = {1461-1464},
year = {2020},
doi = {10.1126/science.aaz0242},
URL = {https://www.science.org/doi/abs/10.1126/science.aaz0242}
}

@article{ASYP21,
  title = {Dynamical Fermionization in One-Dimensional Spinor Quantum Gases},
  author = {Alam, Shah Saad and Skaras, Timothy and Yang, Li and Pu, Han},
  journal = {Phys. Rev. Lett.},
  volume = {127},
  issue = {2},
  pages = {023002},
  numpages = {6},
  year = {2021},
  month = {Jul},
  publisher = {American Physical Society},
  doi = {10.1103/PhysRevLett.127.023002},
  url = {https://link.aps.org/doi/10.1103/PhysRevLett.127.023002}
}

@article{Patu23,
  title = {{Dynamical Fermionization in One-Dimensional Spinor Gases at Finite Temperature}},
  author = {P\^{a}\c{t}u, Ovidiu I.},
  journal = {Phys. Rev. Lett.},
  volume = {130},
  issue = {16},
  pages = {163201},
  numpages = {7},
  year = {2023},
  month = {Apr},
  publisher = {American Physical Society},
  doi = {10.1103/PhysRevLett.130.163201},
  url = {https://link.aps.org/doi/10.1103/PhysRevLett.130.163201}
}

@article{GP08,
  title = {{Correlations in an expanding gas of hard-core bosons}},
  author = {Gangardt, D. M. and Pustilnik, M.},
  journal = {Phys. Rev. A},
  volume = {77},
  issue = {4},
  pages = {041604},
  numpages = {4},
  year = {2008},
  month = {Apr},
  publisher = {American Physical Society},
  doi = {10.1103/PhysRevA.77.041604},
  url = {https://link.aps.org/doi/10.1103/PhysRevA.77.041604}
}

@article{BHLM12,
  title = {{Long-Time Behavior of the Momentum Distribution During the Sudden Expansion of a Spin-Imbalanced Fermi Gas in One Dimension}},
  author = {Bolech, C. J. and Heidrich-Meisner, F. and Langer, S. and McCulloch, I. P. and Orso, G. and Rigol, M.},
  journal = {Phys. Rev. Lett.},
  volume = {109},
  issue = {11},
  pages = {110602},
  numpages = {5},
  year = {2012},
  month = {Sep},
  publisher = {American Physical Society},
  doi = {10.1103/PhysRevLett.109.110602},
  url = {https://link.aps.org/doi/10.1103/PhysRevLett.109.110602}
}

@article{CGK15,
  title = {{Sudden Expansion of a One-Dimensional Bose Gas from Power-Law Traps}},
  author = {Campbell, A. S. and Gangardt, D. M. and Kheruntsyan, K. V.},
  journal = {Phys. Rev. Lett.},
  volume = {114},
  issue = {12},
  pages = {125302},
  numpages = {6},
  year = {2015},
  month = {Mar},
  publisher = {American Physical Society},
  doi = {10.1103/PhysRevLett.114.125302},
  url = {https://link.aps.org/doi/10.1103/PhysRevLett.114.125302}
}

@article{XR17,
  title = {{Expansion of one-dimensional lattice hard-core bosons at finite temperature}},
  author = {Xu, Wei and Rigol, Marcos},
  journal = {Phys. Rev. A},
  volume = {95},
  issue = {3},
  pages = {033617},
  numpages = {8},
  year = {2017},
  month = {Mar},
  publisher = {American Physical Society},
  doi = {10.1103/PhysRevA.95.033617},
  url = {https://link.aps.org/doi/10.1103/PhysRevA.95.033617}
}

@article{CDDK19,
	title={{Hydrodynamics of the interacting Bose gas in the Quantum Newton Cradle setup}},
	author={Jean-S\'{e}bastien Caux and Benjamin Doyon and J\'{e}r\^{o}me Dubail and Robert Konik and Takato Yoshimura},
	journal={SciPost Phys.},
	volume={6},
	pages={070},
	year={2019},
	publisher={SciPost},
	doi={10.21468/SciPostPhys.6.6.070},
	url={https://scipost.org/10.21468/SciPostPhys.6.6.070},
}

@article{ABGK17,
  title = {{Collective many-body bounce in the breathing-mode oscillations of a Tonks-Girardeau gas}},
  author = {Atas, Y. Y. and Bouchoule, I. and Gangardt, D. M. and Kheruntsyan, K. V.},
  journal = {Phys. Rev. A},
  volume = {96},
  issue = {4},
  pages = {041605},
  numpages = {6},
  year = {2017},
  month = {Oct},
  publisher = {American Physical Society},
  doi = {10.1103/PhysRevA.96.041605},
  url = {https://link.aps.org/doi/10.1103/PhysRevA.96.041605}
}

@article{Patu20,
  title = {{Nonequilibrium dynamics of the anyonic Tonks-Girardeau gas at finite temperature}},
  author = {P\^{a}\c{t}u, Ovidiu I.},
  journal = {Phys. Rev. A},
  volume = {102},
  issue = {4},
  pages = {043303},
  numpages = {21},
  year = {2020},
  month = {Oct},
  publisher = {American Physical Society},
  doi = {10.1103/PhysRevA.102.043303},
  url = {https://link.aps.org/doi/10.1103/PhysRevA.102.043303}
}

@article{FCJB14,
  title = {{Quench-Induced Breathing Mode of One-Dimensional Bose Gases}},
  author = {Fang, Bess and Carleo, Giuseppe and Johnson, Aisling and Bouchoule, Isabelle},
  journal = {Phys. Rev. Lett.},
  volume = {113},
  issue = {3},
  pages = {035301},
  numpages = {5},
  year = {2014},
  month = {Jul},
  publisher = {American Physical Society},
  doi = {10.1103/PhysRevLett.113.035301},
  url = {https://link.aps.org/doi/10.1103/PhysRevLett.113.035301}
}

@article{SSAC08,
  title = {{Reduced density matrices and coherence of trapped interacting bosons}},
  author = {Sakmann, Kaspar and Streltsov, Alexej I. and Alon, Ofir E. and Cederbaum, Lorenz S.},
  journal = {Phys. Rev. A},
  volume = {78},
  issue = {2},
  pages = {023615},
  numpages = {17},
  year = {2008},
  month = {Aug},
  publisher = {American Physical Society},
  doi = {10.1103/PhysRevA.78.023615},
  url = {https://link.aps.org/doi/10.1103/PhysRevA.78.023615}
}

@article{Lode16,
  title = {{Multiconfigurational time-dependent Hartree method for bosons with internal degrees of freedom: Theory and composite fragmentation of multicomponent 
  Bose-Einstein condensates}},
  author = {Lode, Axel U. J.},
  journal = {Phys. Rev. A},
  volume = {93},
  issue = {6},
  pages = {063601},
  numpages = {10},
  year = {2016},
  month = {Jun},
  publisher = {American Physical Society},
  doi = {10.1103/PhysRevA.93.063601},
  url = {https://link.aps.org/doi/10.1103/PhysRevA.93.063601}
}

@article{Yang62,
  title = {{Concept of Off-Diagonal Long-Range Order and the Quantum Phases of Liquid He and of Superconductors}},
  author = {Yang, C. N.},
  journal = {Rev. Mod. Phys.},
  volume = {34},
  issue = {4},
  pages = {694--704},
  numpages = {0},
  year = {1962},
  month = {Oct},
  publisher = {American Physical Society},
  doi = {10.1103/RevModPhys.34.694},
  url = {https://link.aps.org/doi/10.1103/RevModPhys.34.694}
}

@article{GT22a,
  title = {{Stability against three-body clustering in one-dimensional spinless $p$-wave fermions}},
  author = {Guo, Yixin and Tajima, Hiroyuki},
  journal = {Phys. Rev. A},
  volume = {106},
  issue = {4},
  pages = {043310},
  numpages = {10},
  year = {2022},
  month = {Oct},
  publisher = {American Physical Society},
  doi = {10.1103/PhysRevA.106.043310},
  url = {https://link.aps.org/doi/10.1103/PhysRevA.106.043310}
}

@article{GT23,
  title = {{Competition between pairing and tripling in one-dimensional fermions with coexistent $s$- and $p$-wave interactions}},
  author = {Guo, Yixin and Tajima, Hiroyuki},
  journal = {Phys. Rev. B},
  volume = {107},
  issue = {2},
  pages = {024511},
  numpages = {7},
  year = {2023},
  month = {Jan},
  publisher = {American Physical Society},
  doi = {10.1103/PhysRevB.107.024511},
  url = {https://link.aps.org/doi/10.1103/PhysRevB.107.024511}
}

\end{document}